\documentclass[a4paper,11pt]{article}
\pdfoutput=1 
\usepackage{jheppub} 
\usepackage[T1]{fontenc} 
\usepackage{verbatim}

\title{\boldmath A More Natural Composite Higgs Model}

\author{Hsin-Chia Cheng}
\author{and Yi Chung}

\affiliation{Center for Quantum Mathematics and Physics (QMAP), Department of Physics, \\University of California,  Davis, CA 95616, U.S.A.}

\emailAdd{cheng@physics.ucdavis.edu}
\emailAdd{yichung@ucdavis.edu}

\abstract{Composite Higgs models provide an attractive solution to the hierarchy problem. However, many realistic models suffer from tuning problems in the Higgs potential. There are often large contributions from the UV dynamics of the composite resonances to the Higgs potential, and tuning between the quadratic term and the quartic term is required to separate the electroweak breaking scale and the compositeness scale. We consider a composite Higgs model based on the $SU(6)/Sp(6)$ coset, where an enhanced symmetry on the fermion resonances can minimize the Higgs quadratic term. Moreover, a Higgs quartic term from the collective symmetry breaking of the little Higgs mechanism can be realized by the partial compositeness couplings between elementary Standard Model fermions and the composite operators, without introducing new elementary fields beyond the Standard Model and the composite sector. The model contains two Higgs doublets, as well as several additional pseudo-Nambu-Goldstone bosons. To avoid tuning, the extra Higgs bosons are expected to be relatively light and may be probed in the future LHC runs. The deviations of the Higgs couplings and the weak gauge boson couplings also provide important tests as they are expected to be close to the current limits in this model. }

\begin{document} 
\maketitle
\flushbottom

\section{Introduction}\label{sec:Introduction}

The Standard Model (SM) of particle physics successfully describes all known elementary particles and their interactions. At the center of SM is the mechanism of electroweak symmetry breaking (EWSB), which is responsible for the masses of gauge bosons and fermions. The discovery of Higgs bosons in 2012 \cite{Chatrchyan:2012xdj, Aad:2012tfa} filled in the last missing piece of the SM. However, the Higgs boson itself brings new questions and puzzles that need to be answered. As a minimal model to realize EWSB, the Higgs field is characterized by the potential 
\begin{equation}
\label{eq:1}
V(H)=-\mu^2|H|^2+\lambda|H|^4
\end{equation}
with just two parameters. The two parameters are now fixed by the observed Higgs vacuum expectation value (VEV) $v\simeq 246$ GeV and Higgs boson mass $M_h\simeq 125$ GeV as
\begin{equation}
\label{eq:2}
\mu^2\simeq\left(88\text{ GeV}\right)^2,\qquad
\lambda\simeq0.13~.
\end{equation}

However, SM does not address the UV-sensitive nature of scalar bosons.  The Higgs mass-squared receives quadratically divergent radiative corrections from the interactions with SM fields, which leads to the well-known hierarchy problem. To avoid the large quadratic corrections, the most natural way is to invoke some new symmetry such that the quadratic contributions cancel in the symmetric limit. This requires the presence of new particles related to SM particles by the new symmetry, such as top partners, in order to cut off the divergent loop contributions. 

One such appealing solution to the hierarchy problem is the composite Higgs model (CHM), where the Higgs doublet is the pseudo-Nambu-Goldstone boson (pNGB) of a spontaneously broken global symmetry of the underlying strong dynamics~\cite{Kaplan:1983fs, Kaplan:1983sm}. Through the analogy of the chiral symmetry breaking in quantum chromodynamics (QCD), which naturally introduces light scalar fields, i.e., pions, we can construct models with light Higgs bosons in a similar way. In a CHM, an approximate global symmetry $G$ is spontaneously broken by some strong dynamics down to a subgroup $H$ with a symmetry breaking scale $f$. The heavy resonances of the strong dynamics are expected to be around the compositeness scale $\sim 4\pi f$ generically. The pNGBs of the symmetry breaking, on the other hand, can naturally be light with masses $<f$ as they are protected by the shift symmetry. The potential of the Higgs field arises from the explicit symmetry breaking effects, such as the interactions with other SM fields. The largest coupling of the Higgs field in SM is to the top quark. As a result, for naturalness, the top partners which regulate the top loop contribution to the Higgs potential should not be too heavy. The top loop contribution to the Higgs mass term can be estimated as
\begin{equation}
\label{eq:toploop}
\Delta \mu^2\sim \frac{N_c}{8\pi^2}y_t^2M_T^2 \sim
\left(220\text{ GeV}\right)^2\left(\frac{M_T}{1.2\text{ TeV}}\right)^2,
\end{equation}
where $M_T$ is the top partner mass. On the other hand, the bounds on the SM colored top partners have reached beyond 1 TeV from the collider searches~\cite{Sirunyan:2018omb, Aaboud:2018pii}. Compared with Eq.(\ref{eq:2}), we see that the models with colored top partners (including both the minimal supersymmetric standard model (MSSM) and the CHM) already require some unavoidable $\mathcal{O}(10\%)$ tuning, albeit not unimaginable.

In most CHMs, however, the tuning is much worse than that is shown in Eq.~(\ref{eq:toploop}). Depending on the coset $G/H$ and the representations of composite operators that couple to the top quarks, the strongly interacting resonances of the top sector in the UV often give a bigger contribution to the Higgs potential than Eq.~(\ref{eq:toploop}), which requires more tuning to cancel. Another problem is that, unlike the pions, the Higgs field needs to develop a nonzero VEV $v$. The current experimental constraints require $v < f/3$. On the other hand, for a generic pNGB potential, the natural VEV for the pNGB is either 0 or $f$. To obtain a VEV much less than $f$, a significant quartic Higgs potential compared to the quadratic term is needed. In little Higgs models~\cite{ArkaniHamed:2001nc, ArkaniHamed:2002qx, ArkaniHamed:2002qy}, a Higgs quartic term can be generated without inducing a large quadratic term from the collective symmetry breaking. Such a mechanism is not present in most CHMs, which is another cause of the fine-tuning issue.

In this study, our goal is to find a more natural CHM by removing the additional tuning beyond Eq.~(\ref{eq:toploop}). We first identify the cosets and the composite operator representations that couple to the top quarks, which can preserve a larger symmetry for the resonances to suppress the UV contribution to the Higgs potential. Next, we implement the collective symmetry breaking to generate a Higgs quartic potential while keeping the quadratic term at the level of Eq.~(\ref{eq:toploop}). In this way we can naturally separate the scales of $v$ and $f$, resulting in a more natural CHM.

This paper is organized as follows. In section~\ref{sec:Motivation}, we review the tuning problems in CHMs and identify the sources of the extra tuning, using the $SO(5)/SO(4)$ CHMs as an example. In section~\ref{sec:Model}, we introduce the $SU(6)/Sp(6)$ CHM, including the interactions that produce the SM Yukawa couplings, and show how the large UV contribution to the Higgs potential is avoided. We then move on to the next step to generate an independent Higgs quartic term from collective symmetry breaking in section~\ref{sec:Quartic}. The resulting Higgs potential of the 2HDM is discussed in section~\ref{sec:2HDM}. The complete potential and spectrum of all the pNGBs in our model are summarized in section~\ref{sec:Spectrum} with numerical estimation. Section~\ref{sec:Collider} and Section~\ref{sec:Precision} are devoted to the phenomenology of this model. Section~\ref{sec:Collider} focuses on the collider searches and constraints. The analyses of the indirect constraints from the precision experimental measurements are presented in Section~\ref{sec:Precision}. Section~\ref{sec:Conclusion} contains our summaries and conclusions. In Appendix~\ref{SU(5)/SO(5)} we briefly discuss the possibility of constructing a similar model based on the $SU(5)/SO(5)$ coset. We point out the differences and some drawbacks of such a model. Appendix~\ref{PQ charge} contains the details of the interactions between elementary fermions and composite operators for a realistic implementation of the $SU(6)/Sp(6)$ CHM model.

\section{Tuning in General Composite Higgs Models}\label{sec:Motivation}

We first give a brief review of the tuning problem of the Higgs potential in general CHMs, which was comprehensively discussed in Ref.~\cite{Panico:2012uw, Panico:2015jxa}. This will help to motivate possible solutions. As an illustration, we consider the Minimal Composite Higgs Models (MCHMs)~\cite{Agashe:2004rs} with the symmetry breaking $SO(5) \to SO(4)$. The four pNGBs are identified as the SM Higgs doublet. The SM gauge group $SU(2)_W\times U(1)_Y$ is embedded in $SO(5)\times U(1)_X$, with the extra $U(1)_X$ accounting for the hypercharges of SM fermions. The explicit breaking of the global symmetry introduces a pNGB potential such that at the minimum the $SO(5)$ breaking VEV $f$ is slightly rotated away from the direction that preserves the $SU(2)_W\times U(1)_Y$ gauge group. The misalignment leads to the EWSB at a scale $v \ll f$.

The explicit global symmetry breaking comes from SM gauge interactions and Yukawa interactions. The SM Yukawa couplings arise from the partial compositeness mechanism~\cite{Kaplan:1991dc}: elementary fermions mix with composite operators of the same SM quantum numbers from the strong dynamics,
\begin{equation}
\mathcal{L}=\lambda_{L}\bar{q}_{L}{O}_{R}+\lambda_{R}\bar{q}_{R}{O}_{L},
\end{equation}
where $q_L$, $q_R$ are elementary fermions and ${O}_L$, ${O}_R$ are composite operators of some representations of $G$ ($=SO(5)$ in MCHMs). The values of couplings $\lambda_{L}$, $\lambda_{R}$ depend on the UV theory of these interactions and are treated as free parameters to produce viable models. With these interactions, the observed SM fermions will be mixtures of elementary fermions and composite resonances. The SM fermions can then couple to the Higgs field through the portion of the strong sector with couplings given by
\begin{equation}
y\simeq\frac{\lambda_{L}\lambda_{R}}{g_\psi}\simeq \epsilon_L\cdot g_\psi\cdot\epsilon_R~,
\end{equation}
where $g_\psi$ is a coupling of the strong resonances and is expected to be $\gg 1$, $\epsilon_{L, R}$ are ratios $\lambda_{L,R}/g_\psi$, which are expected to be small. The resonances created by ${O}_{L,R}$ have masses $\sim g_\psi f$, and play the roles of SM fermion partners. They cut off the divergent contributions to the Higgs potential and make it finite. Notice that the operators belong to representations of the global symmetry $G$, but the resonances are divided into representations of $H$ after the symmetry breaking. Because the elementary fermions in general do not fill the whole representations of $G$, the partial compositeness couplings $\lambda_{L}$, $\lambda_{R}$ explicitly break the global symmetry $G$ and generate a nontrivial Higgs potential.

The pNGB Higgs field parametrizes the coset $G/H$ so the potential is periodic in the Higgs field. The Higgs potential can be expanded in $\sin (H/f)$ and up to the quartic term it takes the form 
\begin{equation}
V(H)=-\hat{\alpha} f^2\text{sin}^2\frac{H}{f}+\hat{\beta} f^2\text{sin}^4\frac{H}{f}\label{eqA}~,
\end{equation}
where $\hat{\alpha}$ and $\hat{\beta}$ have mass dimension two and $\hat{\alpha}$ corresponds to the mass-squared parameter of the Higgs field while $\hat{\beta}/f^2$ will contribute to the quartic term.
By expanding $\sin (H/f)$, higher powers of $H$ can be generated from each term, but for convenience, we will simply call the first term quadratic term and the second term quartic term. The parameters $\hat{\alpha}$ and $\hat{\beta}$ are model dependent and are generated by explicit breaking parameters, like $\lambda_{L}$ and $\lambda_{R}$. Given the potential, we can get the VEV and Higgs mass parameterized as
\begin{equation}\label{mass}
v=\sqrt{\frac{\hat{\alpha}}{2\hat{\beta}}}~f, \qquad M_h^2=8\hat{\beta} \frac{v^2}{f^2}(1-\frac{v^2}{f^2})~.
\end{equation}
The misalignment of the minimum from the SM gauge symmetry preserving direction is parametrized by
\begin{equation}
\xi\equiv\frac{v^2}{f^2}=\text{sin}^2\langle\theta\rangle=\frac{\hat{\alpha}}{2\hat{\beta}} \ll 1~,
\end{equation}
where angle $\langle\theta\rangle\equiv\langle h\rangle /f$. Therefore, for a realistic model, we need $\hat{\alpha}\ll\hat{\beta}$ and at the same time, the correct size of $\hat{\beta}$ to get the observed Higgs boson mass $M_h \simeq 125$ GeV.

From the most explicit symmetry breaking effects of the composite Higgs models, one typically gets $\hat{\alpha} > \hat{\beta}$, which is the source of the tuning problem. For example, in MCHM$_5$~\cite{Agashe:2004rs, Panico:2015jxa}, the SM fermions mix with composite operators $O_L$, $O_R\in \mathbf{5}$ of $SO(5)$. After the symmetry breaking, the composite resonances split into $\mathbf{4}$ and $\mathbf{1}$ representations of $SO(4)$. The mass difference between $\mathbf{4}$ and $\mathbf{1}$ resonances generates a Higgs potential at the compositeness (UV) scale with
\begin{subequations}\label{eq:y}
\begin{align}
\hat{\alpha} \sim \frac{N_c}{16\pi^2}\lambda_{L,R}^2M_\psi^2
\sim\epsilon_{L,R}^2\frac{N_cg_\psi^4}{16\pi^2}f^2,\\
\hat{\beta} \sim \frac{N_c}{16\pi^2}\lambda_{L,R}^4f^2
\sim\epsilon_{L,R}^4\frac{N_cg_\psi^4}{16\pi^2}f^2. 
\end{align}
\end{subequations}
The quartic term coefficient $\hat{\beta}$ arises at a higher order in $\epsilon$ than $\hat{\alpha}$, so generically $\hat{\beta} \ll \hat{\alpha}$ is expected instead. It is then required more fine-tuning to achieve the correct EWSB. In some models, it is possible to have $\hat{\alpha} \sim \hat{\beta}$. For example, MCHM$_{14}$ \cite{Panico:2012uw} with  $O_L$, $O_R\in \mathbf{14}$ of $SO(5)$ can lead to the potential with
\begin{align}
\hat{\alpha} \sim \hat{\beta} \sim \frac{N_c}{16\pi^2}\lambda_{L,R}^2 M_\psi^2
\sim\epsilon_{L,R}^2\frac{N_cg_\psi^4}{16\pi^2}f^2,
\label{UV_MCHM14}
\end{align}
where $\hat{\beta}$ arises at the same order as $\hat{\alpha}$. It requires less tuning to achieve $\xi \ll 1$. This has been called ``minimal tuning.'' But even so, the UV contribution of Eq.~\eqref{UV_MCHM14} to $\hat{\alpha}$ is larger than the IR contribution from the top quark loop
\begin{align}
\Delta m^2_{\text{IR}} \sim \frac{N_c}{16\pi^2}y_t^2M_T^2
\sim\epsilon_{L,R}^4\frac{N_cg_\psi^4}{16\pi^2}f^2,
\end{align}
which already requires some levels of fine-tuning as shown in Eq. \eqref{eq:toploop}. This additional UV contribution actually worsens the condition and requires more tuning. A less-tuned scenario is to have a composite right-handed top quark (which is a singlet of $G$). In this case, $\epsilon_R \sim 1$ but does not contribute to the Higgs potential. The Higgs potential is controlled by $\lambda_L \sim y_t$, which can be smaller.

From the above discussion, one can see that to obtain a more natural Higgs potential in CHM, it would be desirable to suppress the contribution from the composite top-partner resonances to the quadratic term. For example, a maximal symmetry was proposed in Ref.~\cite{Csaki:2017cep} to keep the degeneracy of the whole $G$ representation of the top-partner resonances. However, the maximal symmetry is somewhat ad hoc within a simple model and its natural realization requires more complicated model constructions by doubling the global symmetry groups or invoking a holographic extra dimension~\cite{Csaki:2018zzf, Blasi:2020ktl}. We will look for cosets $G/H$ such that the representation of the top-partner resonances do not split even after the symmetry breaking of $G \to H$ so that it preserves a global symmetry $G$ in any single partial compositeness coupling to prevent unwanted large contributions to the Higgs potential. Besides, we need some additional contribution to the quartic term without inducing the corresponding quadratic term simultaneously to make $\hat{\beta} >\hat{\alpha}$ naturally. This may be achieved by the collective symmetry breaking of the little Higgs mechanism~\cite{ArkaniHamed:2001nc, ArkaniHamed:2002qx, ArkaniHamed:2002qy}. Previous attempts include adding exotic elementary fermions to an $SU(5)/SO(5)$ CHM model~\cite{Vecchi:2013bja} and a holographic model with double copies of the global symmetry~\cite{Csaki:2017eio}. Another way of generating the quartic term without the quadratic term using the Higgs dependent kinetic mixing requires both new elementary fermions and an enlarged global symmetry or an extra dimension~\cite{Csaki:2019coc}. We will take a more economical approach by implementing the little Higgs mechanism without adding exotic elementary fermions or invoking multiple copies of the global symmetry, but simply using the couplings that mix SM fermions with composite resonances.

\section{The $SU(6)/Sp(6)$ Composite Higgs Model}\label{sec:Model}

Among the possible cosets, the cosets $SU(5)/SO(5)$ and $SU(6)/Sp(6)$ are potential candidates to realize the ideas discussed at the end of the previous section. If the composite operator $O_{L,R}\in \mathbf{5}(\mathbf{6})$ of $SU(5)(SU(6))$, the corresponding resonances do not split under the unbroken subgroup $SO(5)(Sp(6))$.\footnote{Na\"ively they can split into two real representations, but if they carry charges under the extra $U(1)_X$ gauge group which is required to obtain the correct hypercharge, they need to remain complex.} Since they are still complete multiplets of $G$, there is an enhanced symmetry for each mixing coupling $\lambda_{L,R}$, which protects the pNGB potential.  The cosets were also some earliest ones employed in little Higgs models~\cite{ArkaniHamed:2002qy,Low:2002ws} where the collective symmetry breaking for the quartic coupling was realized. In CHMs, it requires different explicit implementations if no extension of the SM gauge group or extra elementary fermions are introduced. The $SU(5)/SO(5)$ model has a general problem that an $SU(2)$ triplet scalar VEV violates the custodial $SU(2)$ symmetry, leading to strong experimental constraints. We will focus on the $SU(6)/Sp(6)$ model\footnote{A CHM with the $SU(6)/Sp(6)$ coset were considered in Ref.~\cite{Cai:2018tet}, but for a different prospect.} here and leave a brief discussion of the $SU(5)/SO(5)$ model in Appendix \ref{SU(5)/SO(5)}.

\subsection{Basics of $SU(6)/Sp(6)$}
To parametrize the $SU(6)/Sp(6)$ non-linear sigma model, we can use a sigma field $\Sigma^{ij}$, which transforms as an anti-symmetric tensor representation $\mathbf{15}$ of $SU(6)$, where $i, j=1, \ldots 6$ are $SU(6)$ indices. The transformation can be expressed as $\Sigma \to g\Sigma g^T$ with $g\in SU(6)$ or as $\Sigma^{ij} \to {g^i}_k{g^j}_\ell\Sigma^{k\ell}$ with indices explicitly written out. The scalar field $\Sigma$ has an anti-symmetric VEV $\langle \Sigma\rangle=  \Sigma_0^{\alpha\beta}$ (with $\alpha$, $\beta$ representing $Sp(6)$ index), where
\begin{equation}
\Sigma_0=
\begin{pmatrix}
0  &  -\mathbb{I} \\
\mathbb{I}   &  0 \\
\end{pmatrix},
\end{equation}
and $\mathbb{I}$ is the $3\times3$ identity matrix. The $\Sigma$ VEV breaks $SU(6)$ down to $Sp(6)$, producing 14 Nambu-Goldstone bosons.

The 35 $SU(6)$ generators can be divided into the unbroken ones and broken ones with each type satisfying 
\begin{equation}
\begin{cases}
\text{unbroken generators}   &T_a   : T_a\Sigma_0+\Sigma_0T_a^T=0~,\\
\text{broken generators}        &X_a   : X_a\Sigma_0-\Sigma_0X_a^T=0~.
\end{cases}
\end{equation}
The Nambu-Goldstone fields can be written as a matrix with the broken generator:
\begin{equation}
\xi(x)={\xi^i}_\alpha(x)\equiv e^{\frac{i\pi_a(x)X_a}{2f}}. 
\end{equation}
Under $SU(6)$, the $\xi$ field transforms as $\xi \to g \xi h^{\dagger}$ where $g \in SU(6)$ and $h \in Sp(6)$, so $\xi$ carries one $SU(6)$ index and one $Sp(6)$ index. 
The relation between $\xi$ and $\Sigma$ field is given by
\begin{equation}
\Sigma(x)=\Sigma^{ij}(x)\equiv \xi \Sigma_0\xi^T=e^{\frac{i\pi_a(x)X_a}{2f}} \Sigma_0e^{\frac{i\pi_a(x)X_a^T}{2f}}=e^{\frac{i\pi_a(x)X_a}{f}}\Sigma_0~.
\end{equation}
The complex conjugation raises or lowers the indices. The fundamental representation of $Sp(6)$ is (pseudo-)real and the $Sp(6)$ index can be raised or lowered by $\Sigma_0^{\alpha\beta}$ or $\Sigma_{0,\alpha\beta}$.

The broken generators and the corresponding fields in the matrix can be organized as follows ($\epsilon=i\sigma^2$):
\begin{equation}
\pi_aX_a=
\begin{pmatrix}
 \frac{1}{\sqrt{2}}\phi_a\sigma^a-\frac{\eta}{\sqrt{6}}\bf{1}  &  ~H_2    &  ~\epsilon s     & ~H_1    \\
H_2^\dagger   &  ~\frac{2\eta}{\sqrt{6}}   &   ~-H_1^T  &    ~0 \\  
\epsilon^T s^*  &  ~-H_1^*    &    ~\frac{1}{\sqrt{2}}\phi_a\sigma^{a*}-\frac{\eta}{\sqrt{6}}\bf{1}   &  ~H_2^*\\
H_1^\dagger   &  ~0   &   ~H_2^T  &   ~\frac{2\eta}{\sqrt{6}} \\  
\end{pmatrix}~.
\end{equation}
In this matrix, there are 14 independent fields. They are (under $SU(2)_W$): a real triplet $\phi_a$, a real singlet $\eta$, a complex singlet $s$, and two Higgs (complex) doublets $H_1$ and $H_2$. We effectively end up  with a two-Higgs-doublet model (2HDM). The observed Higgs boson will correspond to a mixture of $h_1$ and $h_2$ inside two Higgs doublets $H_1=H_{1/2}\supset \frac{1}{\sqrt{2}}\bigl(\begin{smallmatrix}0\\h_1\end{smallmatrix}\bigr)$ and $H_2=H_{-1/2}\supset\frac{1}{\sqrt{2}}\bigl(\begin{smallmatrix}h_2\\0\end{smallmatrix}\bigr)$. Using the Nambu-Goldstone matrix, we can construct the low energy effective Lagrangian for the Higgs fields and all the other pNGBs.

\subsection{The Gauge Sector}
The SM electroweak gauge group $SU(2)_W\times U(1)_Y$ is embedded in $SU(6)\times U(1)_X$ with generators given by
\begin{equation}
SU(2)_W: \frac{1}{2}
\begin{pmatrix}
\sigma^a   &  0  &  0  &  0 \\
0   &  0  &  0  &  0\\
0   &  0  &  -\sigma^{a*} &  0\\
0   &  0  &  0  &  0  \\  
\end{pmatrix},\quad
U(1)_Y: \frac{1}{2}
\begin{pmatrix}
0   &  0  &  0   &  0  &  0  &  0  \\
0   &  0  &  0   &  0  &  0  &  0  \\
0   &  0  &  1  &  0   &  0  &  0  \\
0   &  0  &  0   &  0  &  0  &  0  \\
0   &  0  &  0   &  0  &  0  &  0  \\
0   &  0  &  0   &  0  &  0  & -1  \\
\end{pmatrix} + X \mathbf{I}~.
\end{equation}
The extra $U(1)_X$ factor accounts for the different hypercharges of the fermion representations but is not relevant for the bosonic fields. These generators belong to $Sp(6)\times U(1)_X$ and not broken by $\Sigma_0$.
Using the $\Sigma$ field, the Lagrangian for kinetic terms of Higgs boson comes from
\begin{equation}
\mathcal{L}_h=\frac{f^2}{4}\text{tr}\left[(D_{\mu}\Sigma)(D^\mu \Sigma)^\dagger\right]+\cdots ,
\end{equation}
where $D_{\mu}$ is the electroweak covariant derivative. Expanding this, we get
\begin{equation}
\mathcal{L}_h=\frac{1}{2}(\partial _\mu h_1)(\partial ^\mu h_1)+\frac{1}{2}(\partial _\mu h_2)(\partial ^\mu h_2)+\frac{f^2}{2}g_W^2 \left(\text{sin}^2\frac{\sqrt{h_1^2+h_2^2}}{\sqrt{2}f}\right) \left[W^+_\mu W^{-\mu}+\frac{Z_\mu Z^\mu}{2\text{cos}\theta_W}\right]~.
\end{equation}
The non-linear behavior of Higgs boson in CHM is apparent from the dependence of trigonometric functions.

The $W$ boson acquires a mass when $h_1$ and $h_2$ obtain nonzero VEVs $V_1$ and $V_2$ of
\begin{equation}
m_W^2=\frac{f^2}{2}g_W^2 \left(\text{sin}^2\frac{\sqrt{V_1^2+V_2^2}}{\sqrt{2}f}\right)= \frac{1}{4}g_W^2(v_1^2+v_2^2)
=\frac{1}{4}g_W^2v^2,
\end{equation}
where
\begin{equation}
v_i\equiv \sqrt{2}f\frac{V_i}{\sqrt{V_1^2+V_2^2}}\text{sin}\frac{\sqrt{V_1^2+V_2^2}}{\sqrt{2}f}\approx V_i=\langle h_i\rangle~.
\end{equation}
The parameter that parametrizes the nonlinearity of the CHM is given by
\begin{equation}
\xi\equiv \frac{v^2}{f^2}= 2\,\sin^2\frac{\sqrt{V_1^2+V_2^2}}{\sqrt{2}f}~.
\end{equation}

\subsection{The Gauge Contribution to the pNGB Potential}

SM gauge interactions explicitly break the $SU(6)$ global symmetry, so they contribute to the potential of the Higgs fields as well as other pNGBs. SM gauge bosons couple to pNGBs through the mixing with composite resonances:
\begin{equation}
\mathcal{L}=gW_{\mu,a}J^{\mu,a}_W+g'B_{\mu}J^{\mu}_Y~.
\end{equation}
The $J_W$ and $J_Y$ belong to the composite operators in an adjoint representation $\mathbf{35}$ of $SU(6)$. After the symmetry breaking, the composite operators are decomposed into $\mathbf{21}$ and $\mathbf{14}$ of $Sp(6)$. The masses of composite resonances of different representations of $Sp(6)$ are in general different and this will generate a potential for pNGBs at $\mathcal{O}(g^2)$. For $SU(2)_W$, it only breaks the global symmetry partially and generates mass terms for the two Higgs doublets and the scalar triplet $\phi$:
\begin{align}
SU(2)_W: ~\text{(for $H_1, H_2$)}~&c_w\frac{1}{16\pi^2}\frac{3g^2}{2}g_\rho^2f^2 \approx c_w\frac{3}{32\pi^2}g^2M_\rho^2 ,\\
\text{(for $\phi$)}~&c_w\frac{1}{16\pi^2}4g^2g_\rho^2f^2 \approx c_w\frac{1}{4\pi^2}g^2M_\rho^2 \label{mphi} ,
\end{align}
where $g_\rho f\sim M_\rho$ is the mass of the vector resonances $\rho$ which act as the gauge boson partners to cut off the $SU(2)_W$ gauge loop contribution to the pNGB masses, and $c_w$ is a $\mathcal{O}(1)$ constant. Similarly, for $U(1)_Y$, the interaction also breaks the global symmetry partially. It only generates mass terms for $H_1$, $H_2$:
\begin{align}
U(1)_Y: ~c'\frac{1}{32\pi^2}g'^2g_{\rho}^2f^2\approx c'\frac{1}{32\pi^2}g'^2M_{\rho}^2,
\end{align}
where $c'$ is also an $\mathcal{O}(1)$ constant.

Combining these two contributions, we get the mass terms of the pNGBs from the gauge contributions at the leading order as
\begin{align}\label{pNGBmass}
&M_\eta^2=M_s^2=0,\qquad\qquad
M_\phi^2=c_w\frac{1}{4\pi^2}g^2M_\rho^2~,\notag\\
&M_{H_1}^2=M_{H_2}^2=c_w\frac{3}{32\pi^2}g^2M_\rho^2+c'\frac{1}{32\pi^2}g'^2M_{\rho}^2\approx\left(\frac{3g^2+g'^2 (c'/c_w)}{8g^2}\right)M_\phi^2~.
\end{align}
From the gauge contributions only, we expect that $M_\phi>M_{H_1}=M_{H_2}$ and they are below the symmetry breaking scale $f$. The $SU(2)_W \times U(1)_Y$ singlets $s$ and $\eta$ do not receive masses from the gauge interactions at this order, but they will obtain masses elsewhere which will be discussed later.

\subsection{The Yukawa Sector}\label{Yukawa}

For partial compositeness, the elementary quarks and leptons couple to composite operators of $G=SU(6)$. To be able to mix with the elementary fermions, the representations of the composite operators must contain states with the same SM quantum numbers as the SM fermions. For our purpose, we can consider $\mathbf{6}$ and $\mathbf{\bar{6}}$ of $SU(6)$ as they don't split under the $Sp(6)$ subgroup. To account for the correct hypercharge, e.g., $q_L=2_{1/6}$, $q_R=1_{2/3}$ for up-type quarks and $q_R=1_{-1/3}$ for down-type quarks, the composite operators need to carry additional charges under the $U(1)_X$ outside $SU(6)$ and the SM hypercharge is a linear combination of the $SU(6)$ generator $\text{diag}(0,0,1/2, 0,0, -1/2)$ and $X$. The composite operator as a $\mathbf{6}_{1/6}$ of $SU(6)$ (where the subscript $1/6$ denotes its $U(1)_X$ charge) can be decomposed under SM $SU(2)_W \times U(1)_Y$ gauge group as
\begin{equation}
O_{L,R}^i\sim{\xi^i}_\alpha Q_{L,R}^\alpha\sim
\mathbf{6}_{1/6}=\mathbf{2}_{1/6}\oplus\mathbf{1}_{2/3}\oplus\mathbf{\bar{2}}_{1/6}\oplus\mathbf{1}_{-1/3},
\end{equation}
where $Q_{L,R}$ are the corresponding composite resonances. The composite states $Q_{L,R}$ created by these operators belong to the $\mathbf{6}$ representations of $Sp(6)$ and play the roles of SM fermion composite partners. For $SU(2)$, $\mathbf{2}$ and $\mathbf{\bar{2}}$ are equivalent and related by the $\epsilon$ tensor. We make the distinction to keep track of the order of the fermions in a doublet. We see that the composite states have the appropriate quantum numbers to mix with the SM quarks.

The left-handed elementary top quark can mix with either the first two components or the 4th and 5th components of the sextet. If we assume that it couples to the first two components, the mixing term can be expressed as
\begin{equation}
\lambda_{L}\bar{q}_{La}{\Lambda^a}_iO_R^i=
\lambda_{L}\bar{q}_{La}{\Lambda^a}_i\left({\xi^i}_\alpha Q_{R}^\alpha \right) 
\label{eq:left_partial}
\end{equation}
where $a$ represents an $SU(2)_W$ index, and 
\begin{equation}
{(\Lambda)^a}_i=\Lambda=
\begin{pmatrix}
1   &  0  &  0  & 0  & 0  & 0  \\
0   &  1  &  0  & 0  & 0  & 0  \\
\end{pmatrix}
\end{equation}
is the spurion which keeps track of the symmetry breaking.

To get the top Yukawa coupling, we couple the elementary right-handed quark to the $\mathbf{\bar{6}}_{1/6}$, which decomposes under $SU(2)_W \times U(1)_Y$ as
\begin{equation}
O'_{L,Rj}\sim{\xi^*_j}^\beta \Sigma_{0\beta\alpha}Q_{L,R}^\alpha\sim
\mathbf{\bar{6}}_{1/6}=\mathbf{\bar{2}}_{1/6}\oplus\mathbf{1}_{-1/3}\oplus\mathbf{2}_{1/6}\oplus\mathbf{1}_{2/3}~.
\end{equation}
The right-handed top quark mixes with the last component of the $\mathbf{\bar{6}}_{1/6}$, which can be written as
\begin{equation}
\lambda_{t_R}\bar{t}_{R}{\Gamma_{t_R}}^jO'_{Lj}=
\lambda_{t_R}\bar{t}_{R}{\Gamma_{t_R}}^j\left({\xi^*_j}^\beta \Sigma_{0\beta\alpha}Q_{L}^\alpha \right),
\label{eq:right_partial}
\end{equation}
where $\Gamma_{t_R}=\left(0~0~0~0~0~1 \right)$ is the corresponding spurion.

Combining $\lambda_{L}$ and $\lambda_{t_R}$ couplings, we can generate the SM Yukawa coupling for the top quark (and similarly for other up-type quarks),\footnote{If we had coupled the left-handed quarks to the 4th and 5th components of $O_R$, 
\begin{equation*}
\tilde{\lambda}_{L}\bar{q}_{La}\epsilon^{ab}{{\Lambda'}_{bi}}O_R^i=
\tilde{\lambda}_{L}\bar{q}_{La}\epsilon^{ab}{{\Lambda'}_{bi}}\left({\xi^i}_\alpha Q_{R}^\alpha \right) + \text{h.c.},
\end{equation*}
with the spurion
\begin{equation*}
{(\Lambda')}_{bi}={\Lambda'}=
\begin{pmatrix}
0   &  0  &  0  & 1  & 0  & 0  \\
0   &  0  &  0  & 0  & 1  & 0  \\
\end{pmatrix}.
\end{equation*}
The combination of $\tilde{\lambda}_L$ and $\lambda_{t_R}$ would generate an up-type Yukawa coupling with $H_1$, $\sim \tilde{\lambda}_{L}\lambda_{t_R}\left(\bar{q}_L\tilde{H}_1 t_R\right)$.}
\begin{equation}
\sim 
\lambda_{L}\lambda_{t_R}\bar{q}_{La}{\Lambda^a}_i{\xi^i}_\alpha \Sigma_{0}^{\alpha\beta}{\xi^T_\beta}^j{\Gamma}^\dagger_{t_Rj} t_R
=\lambda_{L}\lambda_{t_R}\bar{q}_{La}{\Lambda^a}_i \Sigma^{ij}{\Gamma}^\dagger_{t_Rj} t_R\supset \lambda_{L}\lambda_{t_R}\left(\bar{q}_LH_2t_R\right)~.
\label{eq:top_yukawa}
\end{equation}

Similarly, for the bottom quark (or in general down-type quarks), we can couple $b_R$ to the third component of  $\mathbf{\bar{6}}_{1/6}$ with the coupling $\lambda_{b_R}$ and  spurion $\Gamma_{b_R}=\left(0~0~1~0~0~0 \right)$. This generates a bottom Yukawa coupling of
\begin{equation}
\sim 
\lambda_{L}\lambda_{b_R}\bar{q}_{La}{\Lambda^a}_i{\xi^i}_\alpha \Sigma_{0}^{\alpha\beta}{\xi^T_\beta}^j{\Gamma}^\dagger_{b_Rj} b_R
=\lambda_{L}\lambda_{b_R}\bar{q}_{La}{\Lambda^a}_i \Sigma^{ij}{\Gamma}^\dagger_{b_Rj} b_R\supset \lambda_{L}\lambda_{b_R}\left(\bar{q}_LH_1b_R\right)~.
\end{equation}

Alternatively, we could also couple the left-handed elementary quarks to $\mathbf{\bar{6}}_{1/6}$ and right-handed elementary quarks to $\mathbf{6}_{1/6}$,
\begin{equation}
\lambda'_{L}\bar{q}_{La}\epsilon^{ab}{\Omega_b}^iO'_{Ri}=
\lambda'_{L}\bar{q}_{La}\epsilon^{ab}{\Omega_b}^i\left({\xi^*_i}^\beta \Sigma_{0\beta\alpha}Q_{R}^\alpha \right),
\end{equation}
where
\begin{equation}
{(\Omega)_a}^i=\Omega=
\begin{pmatrix}
1   &  0  &  0  & 0  & 0  & 0  \\
0   &  1  &  0  & 0  & 0  & 0  \\
\end{pmatrix}
\end{equation}
and 
\begin{equation}
\lambda_{b_R}'\bar{b}_{R}{\Gamma_{b_R}'}_jO_L^j=
\lambda_{b_R}'\bar{b}_{R}{\Gamma_{b_R}'}_j\left({\xi^j}_\alpha Q_{L}^\alpha \right),
\end{equation}
where $\Gamma_{b_R}'=\left(0~0~0~0~0~1 \right)$.
Combining $\lambda_{L}'$ and $\lambda_{b_R}'$ coupling, we can generate the SM Yukawa coupling for bottom quark as
\begin{equation}
\sim 
\lambda_{L}'\lambda_{b_R}'\bar{q}_{La}\epsilon^{ab}{\Omega_b}^i{\xi^*_i}^\beta \Sigma_{0\beta\alpha}{\xi^{\dagger\alpha}}_j\Gamma_{b_R}'^{*j} b_R
=\lambda_{L}'\lambda_{b_R}'\bar{q}_{La}\epsilon^{ab}{\Omega_b}^i\Sigma^\dagger_{ij}\Gamma_{b_R}'^{*j} b_R
\supset \lambda_{L}'\lambda_{b_R}'\left(\bar{q}_L\tilde{H}_2b_R\right)~,
\end{equation}
where $\tilde{H}\equiv \epsilon H^*$. In this case, the bottom mass also comes from VEV of $H_2$. Note that the combination of $\lambda_L$ and $\lambda_{b_R}'$ (or $\lambda_{L}'$ and $\lambda_{b_R}$) does not generate the SM Yukawa coupling because it does not depend on $\Sigma$.

The lepton Yukawa couplings can be similarly constructed by coupling elementary leptons to $\mathbf{6}$ and $\mathbf{\bar{6}}$ with $X=-1/2$. In 2HDMs, if the SM quarks have general couplings to both Higgs doublets, large tree-level flavor-changing effects can be induced. To avoid them, it is favorable to impose the natural flavor conservation~\cite{Glashow:1976nt,Paschos:1976ay} such that all up-type quarks couple to one Higgs doublet and all down-type quarks couple to either the same Higgs doublet (Type-I) or the other Higgs doublet (Type-II or flipped depending on the lepton assignment). We can obtain all different possibilities by choosing the partial compositeness couplings. For Type-II and flipped models, the $b \to s \gamma$ put strong constraints on the charged Higgs boson mass ($\gtrsim 600$~GeV) \cite{Misiak:2017bgg} which would require more tuning in the Higgs potential. Therefore, we will assume the Type-I 2HDM for the remaining of the paper, with the top Yukawa coupling coming from $\lambda_{L}\lambda_{t_R}$ and the bottom Yukawa coupling coming from $\lambda_{L}'\lambda_{b_R}'$.

\subsection{The Top Contribution to the pNGB Potential}

The partial compositeness coupling $\lambda_{L}$ or $\lambda_{R}$ individually cannot generate a potential for the pNGBs by itself, because the coupling Eq.~\eqref{eq:left_partial} [or \eqref{eq:right_partial}] preserves an $SU(6)$ symmetry represented by the $\alpha$ index. Although $\alpha$ is an $Sp(6)$ index, without $\Sigma_0$, it cannot distinguish $Sp(6)$ from $SU(6)$. To generate a nontrivial Higgs potential, we need at least an insertion of $\Sigma_0$, which distinguishes $Sp(6)$ from $SU(6)$. It first arises through the combination of $\lambda_{L}$ and $\lambda_{R}$ in Eq.~\eqref{eq:top_yukawa}, which is just the top Yukawa coupling. Therefore, the first nontrivial Higgs potential shows up at the next order, i.e., $\mathcal{O}(\lambda_{L}^2\lambda_{R}^2)$, as
\begin{equation}\label{H2_mass}
\sim -\frac{N_c}{8 \pi^2} \lambda_{L}^2\lambda_{R}^2 f^4 \left|{(\Lambda)^a}_i{(\Gamma^{*})}_j\Sigma^{ij}\right|^2
\end{equation}
It gives a contribution to the $H_2$ squared-mass term of the order 
\begin{align}\label{eq:alpha}
\Delta M_{H_2}^2
\sim-\frac{N_c}{8\pi^2}\lambda_{L}^2\lambda_{R}^2f^2
\sim-\frac{N_c}{8\pi^2}y_t^2M_T^2~,
\end{align}
which is the same as the IR contribution from the top loop estimated in Eq.~\eqref{eq:toploop}. Therefore, in this model, we avoid the potentially large $\mathcal{O}(\lambda_{}^2)$ UV contribution and achieve the minimal tuning for the quadratic part of the Higgs potential.

\section{Collective Higgs Quartics from Fermion Partial Compositeness Couplings}\label{sec:Quartic}

In the previous section, we show that in the $SU(6)/Sp(6)$ CHM the UV contribution from the strong dynamics to the Higgs potential is suppressed, minimizing the tuning of the quadratic term. However, we need some additional quartic Higgs potential to further reduce the tuning and to obtain a 125 GeV Higgs boson, as the IR contribution from the top quark loop to the Higgs quartic term is not enough. Generating a Higgs quartic coupling without inducing the corresponding quadratic term is the hallmark of the little Higgs mechanism. For example, in the original $SU(6)/Sp(6)$ little Higgs model \cite{Low:2002ws}, a Higgs quartic term from the collective symmetry breaking can be generated by gauging two copies of $SU(2)$, with generators given by 
\begin{equation}
Q_1^a=\frac{1}{2}
\begin{pmatrix}
\sigma^a   &  0  &  0  &  0 \\
0   &  0  &  0  &  0\\
0   &  0  &  0_{2\times 2} &  0\\
0   &  0  &  0  &  0  \\  
\end{pmatrix}
\quad \text{and} \quad
Q_2^a=-\frac{1}{2}
\begin{pmatrix}
0_{2\times 2}   &  0  &  0  &  0 \\
0   &  0  &  0  &  0\\
0   &  0  &  \sigma^{a*} &  0\\
0   &  0  &  0  &  0  \\  
\end{pmatrix}
\end{equation} 
and gauge couplings $g_1$ and $g_2$. The two $SU(2)$'s are broken down to the diagonal $SU(2)_W$ by the $\Sigma$ VEV. The potential for the pNGBs generated by the two gauge couplings takes the form
\begin{equation}
g_1^2f^2\left|s+\frac{i}{2f}\tilde{H_2}^\dagger H_1\right|^2+
g_2^2f^2\left|s-\frac{i}{2f}\tilde{H_2}^\dagger H_1\right|^2.
\end{equation} 
The $g_1^2$ term preserves the $SU(4)$ symmetry of the $3,4,5,6$ entries which contains the shift symmetry of $H_1$ and $H_2$. If only the first term of the potential exists, the $\tilde{H_2}^\dagger H_1$ dependence can be absorbed into $s$ by a field redefinition and the term just corresponds to a mass term for $s$. Similarly, the $g_2^2$ term preserves the $SU(4)$ symmetry of the $1,2,3,6$ entries under which $H_1$ and $H_2$ remain as Nambu-Goldstone bosons, but with a different shift symmetry. The combination of both terms breaks either of the shift symmetries, and a quartic Higgs potential is generated after integrating out the $s$ field, 
\begin{equation}
\lambda\left|\tilde{H_2}^\dagger H_1\right|^2
\quad\text{with}\quad
\lambda=\frac{g_1^2g_2^2}{g_1^2+g_2^2}~.
\end{equation}

The possibility of gauging two copies of $SU(2)$ gauge group is subject to the strong experimental constraints on $W'$ and $Z'$. We would like to generate the quartic Higgs potential without introducing additional elementary fields to the $SU(6)/Sp(6)$ CHM, so we will consider the collective symmetry breaking from the interactions between the elementary fermions and the resonances of the strong dynamics.

From the discussion of the previous section, we see that the elementary quark doublets may couple to composite operators of $SU(6)$ representations $\mathbf{6}$ and/or $\mathbf{\bar{6}}$, and each contains two doublets of the same SM quantum numbers:
\begin{subequations}
\begin{equation}
\mathbf{6}_{1/6}=\mathbf{2}_{1/6}\oplus\mathbf{1}_{2/3}\oplus\mathbf{\bar{2}}_{1/6}\oplus\mathbf{1}_{-1/3},
\end{equation}
\begin{equation}
\mathbf{\bar{6}}_{1/6}=\mathbf{\bar{2}}_{1/6}\oplus\mathbf{1}_{-1/3}\oplus\mathbf{2}_{1/6}\oplus\mathbf{1}_{2/3}~.
\end{equation}
\end{subequations}
Both operators can create the same resonances which belong to $\mathbf{6}$ of the $Sp(6)$ group.

Now consider two elementary quark doublets couple to the first two components of the composite operators of $\mathbf{6}$ and $\mathbf{\bar{6}}$ respectively, while both representations contain the same resonances:
\begin{equation}
\lambda_{L}\bar{q}_{La}{\Lambda^a}_iO_R^i=
\lambda_{L}\bar{q}_{La}{\Lambda^a}_i\left({\xi^i}_\alpha Q_{R}^\alpha \right),
\label{eq:6}
\end{equation}
where
\begin{equation}
{(\Lambda)^a}_i=\Lambda=
\begin{pmatrix}
1   &  0  &  0  & 0  & 0  & 0  \\
0   &  1  &  0  & 0  & 0  & 0  \\
\end{pmatrix}~,
\end{equation}
and 
\begin{equation}
\lambda'_{L}\bar{q}'_{La}\epsilon^{ab}{\Omega_b}^iO'_{Ri}=
\lambda'_{L}\bar{q}'_{La}\epsilon^{ab}{\Omega_b}^i\left({\xi^*_i}^\beta \Sigma_{0\beta\alpha}Q_{R}^\alpha \right),
\label{eq:6bar}
\end{equation}
where
\begin{equation}
{(\Omega)_a}^i=\Omega=
\begin{pmatrix}
1   &  0  &  0  & 0  & 0  & 0  \\
0   &  1  &  0  & 0  & 0  & 0  \\
\end{pmatrix}~.
\end{equation}
The combination of the two interactions breaks the $SU(6)$ global symmetry explicitly but preserves an $SU(4)$ symmetry of the $3,4,5,6$ entries. It leads to a potential for the pNGBs at $\mathcal{O}(\lambda_{L}^2\lambda_{L}'^{2})$ of the form
\begin{equation}
[{(\Lambda)^a}_i{(\Omega^{*})^b}_j\Sigma^{ij}]
[{{(\Omega)_b}^{m}}{(\Lambda^*)_a}^n\Sigma^{*}_{mn}]~,
\end{equation}
which can easily be checked by drawing a one-loop diagram, with $q_L$, $q'_L$, $Q_R$ running in the loop.

After expanding it we obtain
\begin{equation}
\sim \frac{N_c}{8\pi^2}\lambda_{L}^2\lambda_{L}'^{2} f^4
\left|{(\Lambda)^a}_i{(\Omega^{*})^b}_j\Sigma^{ij}\right|^2
\quad\to\quad
\frac{N_c}{4\pi^2}\lambda_{L}^2\lambda_{L}'^2f^2\left|s+\frac{i}{2f}\tilde{H_2}^\dagger H_1\right|^2~.
\label{eq:collect_1}
\end{equation}
(The factor of 2 comes from the trace which reflects the degrees of freedom running in the loop, as both elementary fermions are doublets.)
This is one of the terms needed for the collective symmetry breaking. The coefficient is estimated from the dimensional analysis.
Notice that we have chosen different (generations of) elementary quark doublets, $q_L$ and $q_L'$ in the two couplings. If $q_L$ and $q_L'$ were the same, the loop can be closed at $\mathcal{O}(\lambda_{L}\lambda_{L}')$ and a large $s$ tadpole term and Higgs quadratic term will be generated,
\begin{equation}\label{s-tadpole}
\sim \frac{N_c}{8\pi^2}\lambda_{L}\lambda_{L}' f^4
\left(\epsilon_{ab}{(\Lambda)^a}_i{(\Omega^{*})^b}_j\Sigma^{ij}\right)
\quad\to\quad
\frac{N_c}{4\pi^2}\lambda_{L}\lambda_{L}' g_{\psi}^2 f^3 \left(s+\frac{i}{2f}\tilde{H_2}^\dagger H_1\right)~.
\end{equation}
Such a term is actually needed for a realistic EWSB, but it would be too large if it were generated together with Eq.~\eqref{eq:collect_1} that will produce the Higgs quartic term. It can be generated of an appropriate size in a similar way involving some other different fermions and composite operators with smaller couplings.

The way that the mass term for $s$ can be generated without the tadpole term can be understood from the symmetry point of view. In addition to the $SU(2)_W \times U(1)_Y$, the $\Sigma_0$ preserves a  global $U(1)$ Peccei-Quinn (PQ)~\cite{Peccei:1977ur} subgroup of $Sp(6)$. This global $U(1)$ symmetry corresponds to the unbroken generator
\begin{equation}
U(1)_{PQ}: \frac{1}{2}
\begin{pmatrix}
1   &  0  &  0   &  0  &  0  &  0  \\
0   &  1  &  0   &  0  &  0  &  0  \\
0   &  0  &  0   &  0   &  0  &  0  \\
0   &  0  &  0   &  -1  &  0  &  0  \\
0   &  0  &  0   &  0  &  -1  &  0  \\
0   &  0  &  0   &  0  &  0  &  0  \\
\end{pmatrix},
\end{equation}
under which $s$ has charge 1, both $H_1, H_2$ have charge 1/2, and the rest of pNGBs have charge 0. The $s$ mass term is invariant under $U(1)_{PQ}$ while the tadpole term has charge 1 so it will not be induced if the interactions can preserve the $U(1)_{PQ}$ symmetry.
On the other hand, the composite operators in Eqs.~\eqref{eq:6}, \eqref{eq:6bar} have the following PQ charges for their components (assuming that they don't carry an additional overall charge),
\begin{subequations}
\begin{align}
&\mathbf{6}_{0}=\mathbf{2}_{1/2}\oplus\mathbf{1}_{0}\oplus\mathbf{\bar{2}}_{-1/2}\oplus\mathbf{1}_{0},\\
&\mathbf{\bar{6}}_{0}=\mathbf{\bar{2}}_{-1/2}\oplus\mathbf{1}_{0}\oplus\mathbf{2}_{1/2}\oplus\mathbf{1}_{0},
\end{align}
\end{subequations}
where the subscript here denotes the PQ charge instead of the $X$ charge.
We see that $q_L$ and $q_L'$ couple to components of different PQ charges. If $q_L$ and $q_L'$ are different, it is possible to assign PQ charges, i.e., $1/2$ for $q_L$ and $-1/2$ for $q_L'$, so that the interactions Eqs.~\eqref{eq:6}, \eqref{eq:6bar} preserve the PQ symmetry and the $s$ tadpole term will not be generated. If $q_L$ and $q_L'$ are the same, then there is no consistent charge assignment that can preserve the PQ symmetry, and hence the $s$ tadpole term can be induced.  Furthermore, if different generations of quarks carry different PQ charges, The $U(1)_{PQ}$ preserving interactions will not induce flavor-changing neutral currents (FCNC) as they violate the PQ symmetry.

The second term required in realizing the collective symmetry breaking can be generated similarly by a different set of quarks (or leptons). They should couple to the 4th and 5th components of the $\mathbf{6}$ and $\mathbf{\bar{6}}$ operators through the spurions
\begin{equation}
{(\Lambda')}_{ai}=
\begin{pmatrix}
0   &  0  &  0  & 1  & 0  & 0  \\
0   &  0  &  0  & 0  & 1  & 0  \\
\end{pmatrix}
\quad \text{and} \quad 
{(\Omega')}^{ai}=
\begin{pmatrix}
0   &  0  &  0  & 1  & 0  & 0  \\
0   &  0  &  0  & 0  & 1  & 0  \\
\end{pmatrix},
\end{equation}
which preserve the $SU(4)$ symmetry of the 1,2,3,6 entries.

The combination of $\Lambda'$ and $\Omega'$ can then introduce the potential 
\begin{equation}
\sim\frac{N_c}{8\pi^2}\lambda_{L}^2\lambda_{L}'^{2} f^4
\left|{(\Lambda')}_{ai}{(\Omega'^{*})}_{bj}\Sigma^{ij}\right|^2
\quad\to\quad
\frac{N_c}{4\pi^2}\lambda_{L}^2\lambda_{L}'^2f^2\left|s-\frac{i}{2f}\tilde{H_2}^\dagger H_1\right|^2,
\end{equation}
which provides the other term needed for the collective symmetry breaking.\\

To generate all the terms required for the Higgs quartic potential from collective symmetry breaking, we need to use several different quarks and/or leptons, with different PQ charge assignments. As we mentioned earlier, we also need some smaller PQ-violating couplings between the elementary fermions and the composite operators, in order to generate a proper-sized  $\tilde{H_2}^\dagger H_1$ term,
\begin{equation}\label{m12}
m_{12}^2\sim \frac{N_c}{8\pi^2}\lambda_{L}\lambda_{L}''g_{\psi}^2f^2,
\end{equation}
where $\lambda_{L}''$ represents the smaller $U(1)_{PQ}$ violating coupling. A more detailed coupling assignment for a realistic model is presented in Appendix \ref{PQ charge}. 
With all the collective symmetry breaking interactions discussed above, we obtain a pNGB potential,

\begin{equation}\label{Vcol}
c_{k\ell}\frac{N_c}{4\pi^2}\lambda_{k_L}^2\lambda_{\ell_L}'^2f^2\left|s+\frac{i}{2f}\tilde{H_2}^\dagger H_1\right|^2+
c_{mn}\frac{N'_c}{4\pi^2}\lambda_{m_L}^2\lambda_{n_L}'^2f^2\left|s-\frac{i}{2f}\tilde{H_2}^\dagger H_1\right|^2 ~,
\end{equation}
where $c_{k\ell}, \, c_{mn}$ are $\mathcal{O} (1)$ constants depending on the UV completion,\footnote{In the Discrete CHMs~\cite{Matsedonskyi:2012ym} or UV completions with Weinberg's sum rules~\cite{Marzocca:2012zn,Pomarol:2012qf} for the MCHM, the analogous finite quartic potentials have the coefficient $c \sim 2$.} and  the indices $k, \ell, m, n$ here label different fermions. After integrating out the massive $s$ field, we obtain a quartic term for the Higgs doublets (take $N_c,N'_c=3$) as
\begin{equation}\label{quartic0} 
\lambda_{12}\left|\tilde{H_2}^\dagger H_1\right|^2
\quad\text{with}\quad
\lambda_{12}=\frac{3}{4\pi^2}\frac{c_{k\ell} c_{mn} \lambda_{k_L}^2\lambda_{\ell_L}'^2\lambda_{m_L}^2\lambda_{n_L}'^2}{c_{k\ell}\lambda_{k_L}^2\lambda_{\ell_L}'^2+ c_{mn}\lambda_{m_L}^2\lambda_{n_L}'^2}~.
\end{equation}
Assuming $ \lambda_{k_L}\lambda_{\ell_L}' \sim\lambda_{m_L}\lambda_{n_L}'$ and $ c_{k\ell} \sim c_{mn}\sim 2$, then in our estimate
\begin{equation}\label{quartic}
\lambda_{12} \sim \frac{3}{4\pi^2}\lambda_{k_L}^2\lambda_{\ell_L}'^2 ~.
\end{equation}

Including this quartic term, the coefficients of the Higgs potential in this model are estimated to be  
\begin{equation}
\hat{\alpha}\sim\frac{3}{16\pi^2}\lambda_{t_L}^2\lambda_{t_R}^2f^2,\quad
\hat{\beta}\sim\frac{3}{16\pi^2}\lambda_{k_L}^2\lambda_{\ell_L}'^2f^2~.
\end{equation}
Therefore we can further improve upon the minimal tuning ($\hat{\alpha}\sim\hat{\beta}$) case by requiring
\begin{equation}
\lambda_{L}'>\lambda_{t_R}\implies \hat{\beta}>\hat{\alpha}~.
\end{equation}
Of course, however, $\hat{\beta}$ can not be arbitrarily large because it is determined by the Higgs boson mass from Eq.~\eqref{mass}. The required numerical parameters will be discussed in the next section.

\section{The Higgs Potential in the 2HDM}\label{sec:2HDM}
The $SU(6)/Sp(6)$ model contains two Higgs doublets.  To analyze the EWSB and the Higgs boson masses, we need to consider the Higgs potential in a 2HDM. A review of 2HDM can be found in Ref.~\cite{Branco:2011iw}. The other pNGBs do not affect the Higgs potential much (they either are heavy or couple mostly quadratically to the Higgs doublets), so we will postpone their discussion to the next section.
The Higgs potential in our model can be parameterized as 
\begin{align}\label{VH}
V(H_1,H_2)&=m_1^2H_1^\dagger H_1+m_2^2H_2^\dagger H_2-m_{12}^2\left(\tilde{H_2}^\dagger H_1+\text{h.c.}\right)\notag\\
&\quad+\frac{\lambda_1}{2}\left(H_1^\dagger H_1\right)^2+\frac{\lambda_2}{2}\left(H_2^\dagger H_2\right)^2+\lambda_{12}\left|\tilde{H_2}^\dagger H_1\right|^2~.
\end{align}
Notice that, in CHMs, due to the non-linearity of pNGBs, the Higgs potential should include trigonometric functions instead of polynomials. Also, to match the potential here to the SM Higgs potential, an additional factor of cos$\langle\theta\rangle$ will appear. However, since the deviation is strongly constrained by Higgs coupling measurements, we will take $\langle\theta\rangle\ll 1$ and expand sin$x\sim x$ in the following discussion for simplicity.

In the 2HDM potential~\eqref{VH}, both Higgs doublets develop nonzero VEVs. Denote the VEVs of $H_1$ and $H_2$ to be $v_1$ and $v_2$ respectively, and their ratio is defined as $\tan \beta\equiv v_2/v_1$. The total VEV $v$ satisfies
\begin{equation}
v^2=v_1^2+v_2^2=v^2 \text{cos}^2\beta+v^2 \text{sin}^2\beta=(246\text{ GeV})^2~.
\end{equation}
$H_2$ couples to the top quark  and gets a large negative loop-induced contribution to its quadratic term, so it is natural to expect $v_2>v_1$. On the other hand, the main quartic term coming from the collective symmetry breaking is $\lambda_{12}$. To have a large enough effective quartic term for the 125 GeV Higgs boson, we do not want either $\sin\beta\, (\equiv s_\beta)$ or $\cos\beta\, (\equiv c_\beta)$ to be too small. The current constraints \cite{Kling:2020hmi, Sirunyan:2019arl, Aaboud:2018cwk} have ruled out the region $\tan\beta$ near 1, so we will consider a benchmark with a medium value,
\begin{equation}
\text{tan}\beta\sim 3~.
\end{equation}
Also, the light neutral eigenstate should be close to the SM Higgs boson, which imposes some conditions on the parameters in the Higgs potential~\eqref{VH}. In Subsec.~\ref{est mass}, we first discuss the quadratic potential, which will determine the spectrum of additional Higgs bosons in this model. Then, we will discuss the alignment issue in Subsec.~\ref{est quartic} and the corresponding values of the quartic terms in the Higgs potential.

\subsection{Estimating the Mass Terms}\label{est mass}

The experimental constraints require that the 2HDM should be close to the alignment limit ($\beta -\alpha=\pi/2$)~\cite{Gunion:2002zf,Craig:2012vn,Craig:2013hca,Carena:2013ooa}, where $\alpha$ is the mixing angle between the mass eigenstates of the two CP-even Higgs boson and the corresponding components in $H_1, H_2$ (after removing the VEVs),
\begin{equation}
h=-h_1\,\sin\alpha +h_2\,\cos\alpha ~.
\end{equation}

To simplify the discussion of the quadratic terms, we assume that the alignment holds approximately,
\begin{equation}
h \approx  h_1\,\cos\beta +h_2\,\sin \beta = h_{\text{SM}},
\end{equation}
then we can calculate the SM Higgs potential by the transformation
\begin{equation}
\left( \begin{array}{c} H_1 \\ H_2 \end{array} \right)= \begin{pmatrix} \text{cos}\beta & -\text{sin}\beta \\ \text{sin}\beta & \text{cos}\beta \end{pmatrix} \left( \begin{array}{c} H_{\text{SM}} \\ H_{\text{heavy}} \end{array} \right) ~.
\end{equation}
The potential of the light SM Higgs doublet becomes (keeping the terms with $H_{\text{SM}}$ only and rewriting $H_{\text{SM}}\to H$)
\begin{align}\label{eq:V_light}
V(H)&=\left(m_1^2\, \text{cos}^2\beta+m_2^2\, \text{sin}^2\beta-2m_{12}^2\, \text{sin}\beta\, \text{cos}\beta\right)|H|^2\notag\\
&\quad+\left(\frac{\lambda_1}{2}\text{cos}^4\beta+\frac{\lambda_2}{2}\text{sin}^4\beta+\lambda_{12}\,\text{sin}^2\beta\, \text{cos}^2\beta\right)|H|^4~.
\end{align}
Matching the quadratic term with the SM Higgs potential implies that
\begin{align}
-\mu^2=m_1^2\, \text{cos}^2\beta+m_2^2\, \text{sin}^2\beta -2m_{12}^2\, \text{sin}\beta\, \text{cos}\beta\approx -\left(88\text{ GeV}\right)^2~.
\label{eq:musquared}
\end{align}
As shown in the previous section, these mass terms get contributions from different sources: $m_1$ comes from gauge contributions, $m_2$ gets an additional large negative contribution from the top quark besides the gauge contributions, and $m_{12}$ comes from the PQ-violating interactions. No natural cancellation among the three terms in Eq.~\eqref{eq:musquared} is warranted. Therefore, the absolute values of all three terms should be of the same order as $\mu^2$ to avoid tuning. 
For example, for $\tan\beta =3$ Eq.~\eqref{eq:musquared} can be satisfied by 
$m_1^2\sim\left(360\text{ GeV}\right)^2$, $m_2^2\sim \left(120\text{ GeV}\right)^2$,
and $m_{12}^2\sim\left(210\text{ GeV}\right)^2$ without strong cancellations among the three terms. These numbers are based on the alignment approximation. More accurate values need to include the whole 2HDM potential and will be given after the discussion of the quartic terms.

\subsection{Estimating the Quartic Terms}\label{est quartic}

There are three quartic couplings in the Higgs potential~\eqref{VH}: $\lambda_1$, $\lambda_2$, and $\lambda_{12}$. The effective quartic coupling for the light Higgs, which can be seen from Eq.~\eqref{eq:V_light}, is a combination of the three quartic couplings and $\tan\beta$. To obtain a 125 GeV Higgs boson we need
\begin{equation}
\frac{\lambda_1}{2}\text{cos}^4\beta+\frac{\lambda_2}{2}\text{sin}^4\beta+\lambda_{12}\,\text{sin}^2\beta\, \text{cos}^2\beta \approx 0.13 ~.
\end{equation}
$\lambda_1$ is mainly induced by the SM gauge loops and is expected to be small. $\lambda_2$ receives the top quark loop contribution,
\begin{equation}\label{eq:toploopquartic}
\lambda_2 \sim \frac{3 y_t^4}{4 \pi^2} \ln \frac{M_T}{v} \sim 0.1 .
\end{equation}
This implies that we need $\lambda_{12}$ which comes from the collective symmetry breaking to satisfy 
\begin{equation}
\lambda_{12}\text{s}_\beta^2 \text{c}_\beta^2 \sim 0.1 ~~\Rightarrow~~ 
\lambda_{12}\sim 1 \quad \text{for } \tan\beta =3 ~.
\end{equation}

If it arises from the collective quartic term obtained in Eq.~\eqref{quartic}, it corresponds to  
\begin{equation}\label{lambda}
\lambda_{L}\lambda_{L}'\sim 3.6 \quad \Rightarrow \quad \sqrt{\lambda_{L}\lambda_{L}'} \sim 1.9 ~.
\end{equation}
These couplings between the elementary states and composite operators are quite large. However, the smallness of SM Yukawa couplings can be obtained by small $\lambda_R$ couplings. There are other experimental constraints with these large $\lambda_L$ couplings, which will be discussed in the following sections.

We have been assuming that the 2HDM potential is approximately in the alignment regime. Let us go back to check how well the alignment can be achieved. A simple way to achieve the alignment is the decoupling limit where the extra Higgs bosons are heavy. However, this would require more tuning in the Higgs mass parameters. In our model $\lambda_{12} > \lambda_2, \lambda_1$. Under this condition, we need $\tan\beta \sim 1$ to achieve the exact alignment if the extra Higgs bosons are not too heavy. This is not compatible with the experiment constraints. Therefore we expect some misalignment and need to check whether the misalignment can be kept within the experimental constraints.

Solving the eigenvalue equations, we can get the following equations for the factor $c_{\beta-\alpha}$,
\begin{align}
c_{\beta-\alpha}&=\frac{1}{M_{A}^2\text{tan}\beta}
\left(\lambda_{1}v_1^2\left(\frac{-s_\alpha}{c_\beta}\right)+\lambda_{12}v_2^2\left(\frac{c_\alpha}{s_\beta}\right)-M_h^2\left(\frac{-s_\alpha}{c_\beta}\right)\right),\\
&=\frac{1}{M_{A}^2\text{cot}\beta}\left(-\lambda_{12}v_1^2\left(\frac{-s_\alpha}{c_\beta}\right)-\lambda_{2}v_2^2\left(\frac{c_\alpha}{s_\beta}\right)+M_h^2\left(\frac{c_\alpha}{s_\beta}\right)\right).
\end{align}
As the misalignment should be small, to estimate its size, we can assume that the mass eigenstates of the 2HDM are near alignment, which satisfy  $(-s_\alpha,c_\alpha)\approx(c_\beta,s_\beta)$ approximately for the right-handed side. We then have
\begin{align}
c_{\beta-\alpha}&\approx\frac{1}{M_{A}^2\text{tan}\beta}
\left(\lambda_{1}v_1^2+\lambda_{12}v_2^2-M_h^2\right),\\
&\approx\frac{1}{M_{A}^2\text{cot}\beta}\left(-\lambda_{12}v_1^2-\lambda_{2}v_2^2+M_h^2\right).
\end{align}
Consider the benchmark values
\begin{equation}
\text{tan}\beta\approx 3,\quad \lambda_{12}\approx 1,\quad\text{and}\quad M_{A}\approx 380\text{ GeV}~,
\end{equation}
where the $M_{A}$ value is chosen to keep the misalignment small and to evade the direct search in the $A^0\to hZ$ decay channel at the  LHC \cite{Kling:2020hmi}. The equations for $c_{\beta-\alpha}$ becomes
\begin{align}\label{misal}
c_{\beta-\alpha}\approx~0.014\lambda_{1}+0.090~\approx~0.199-1.132\lambda_{2}~.
\end{align}
Since $\lambda_{1}$ in this model is small, we have $c_{\beta-\alpha}\approx 0.090$ which parametrizes the deviation from the alignment.
The misalignment will have a direct consequence on Higgs physics and will be discussed in the following sections. The most relevant deviation, the ratio of Higgs to vector bosons coupling to SM coupling, is proportional to $s_{\beta-\alpha}\approx~0.996$ and should still be safe. 

Eq.~\eqref{misal} also implies that $\lambda_{2}$ needs to be $\approx 0.1$, which is consistent with the estimate from the top quark loop contribution Eq.~\eqref{eq:toploopquartic}.
To sum up, the three quartic couplings in our 2HDM potential take values
\begin{equation}
\lambda_{12}\approx 1 \quad\gg\quad\lambda_{2}\approx 0.1 \quad\gg\quad \lambda_{1}~.
\end{equation}

\subsection{A Realistic Higgs Potential}\label{realistic}

So far, all numbers in the above discussion are estimations based on simplified approximations. In a realistic benchmark model, the exact values can be solved by directly diagonalizing the mass matrix. To reproduce the correct Higgs boson mass $M_h=125$ GeV and small enough $c_{\beta-\alpha}$ with fixed tan$\beta\approx3$ and $\lambda_{12}\approx1$, we choose the following values as a reference for our study:
\begin{equation}\label{mA}
\text{tan}\beta\approx3.0,\quad \lambda_{12}\approx1.0,\quad \lambda_{2}\approx0.12,\quad\text{and}\quad M_{A}\approx 380\text{ GeV}~.
\end{equation}
$\lambda_{1}$ is irrelevant as long as it is small so we don't set its value. The value of $\lambda_{2}$ is set by producing the correct Higgs boson mass. 

With these numbers, we can diagonalize the mass matrix and get the mixing angle $\alpha$ and the misalignment $\beta-\alpha$ as
\begin{align}
s_\alpha=-0.215,\quad  c_\alpha=0.977 
\quad\implies\quad c_{\beta-\alpha}=0.1049,\quad s_{\beta-\alpha}=0.9945~.
\end{align}
The eigenvalues of the matrix give the masses of the CP-even neutral scalar bosons as
\begin{align}
M_{h}\approx 125\text{ GeV}\quad\text{and}\quad M_{H}\approx 370\text{ GeV}~.
\end{align}
The complete spectrum will be discussed in the next section.

After we obtain the quartic couplings, we can go back to determine the mass terms. The value of $M_{A}$ is chosen to satisfy the experimental constraint. It also gives the value of $m_{12}$ based on the relation
\begin{align}
m_{12}^2=M_{A}^2{s_\beta c_\beta}\sim\left(210\text{ GeV}\right)^2~.
\end{align}
Given the values of all the quartic couplings and $m_{12}$, we can obtain the other mass terms
\begin{align}
&m_1^2=3m_{12}^2-\frac{1}{2}\lambda_1v_1^2-\frac{1}{2}\lambda_{12}v_2^2
\sim\left(320\text{ GeV}\right)^2,\\
&m_2^2=\frac{1}{3}m_{12}^2-\frac{1}{2}\lambda_2v_2^2-\frac{1}{2}\lambda_{12}v_1^2
\sim \left(90\text{ GeV}\right)^2~.
\end{align}
These numbers will serve as a benchmark for our phenomenological studies.

Assuming that these masses arise dominantly from the loop contributions discussed in the previous sections, we can also estimate the masses of the composite states in the CHM,
\begin{align}
&m_1^2=\frac{3}{32\pi^2}g^2M_\rho^2\sim\left(320\text{ GeV}\right)^2,\\
&m_2^2=\frac{3}{32\pi^2}g^2M_\rho^2-\frac{3}{8\pi^2}y_t^2M_T^2\sim \left(90\text{ GeV}\right)^2,\\
&m_{12}^2=\frac{N_c}{8\pi^2}\lambda_{L}\lambda_{L}''g_{\psi}^2f^2\sim\left(210\text{ GeV}\right)^2~,
\end{align}
where we have ignored the small $U(1)$ gauge contribution and taken $c_w \sim 1$. The $m_1^2$ equation gives the mass of the gauge boson partners $M_\rho\sim 5$ TeV. In the $m_2^2$ equation, the top loop contribution needs to cancel the positive gauge contribution  $\left(320\text{ GeV}\right)^2$  to produce a $\left(90\text{ GeV}\right)^2$ term. From that, the top partner is estimated to be around $M_T\sim 1.6$ TeV. This corresponds to an $\mathcal{O}(10\%)$ tuning between the gauge contribution and the top contribution, but it is hard to avoid given the experimental constraints on the top partner mass. The desired size of $m_{12}^2$ can be achieved by a suitable choice of the PQ-violating coupling $\lambda''_L$ which is a free parameter in this model.

\section{The Spectrum of pNGBs}\label{sec:Spectrum}
After discussing the Higgs potential from the naturalness consideration, we are ready to provide the estimates of  masses of all other pNGBs, based on the benchmark point alluded in the previous section.

\subsection{The Second Higgs Doublet}
The 2HDM potential has been discussed in the previous section. In addition to the SM-like 125 GeV Higgs boson, there is one more CP-even neutral scalar $H^0$, a CP-odd neutral scalar $A^0$, and a complex charge scalar $H^\pm$. Their masses from the Higgs potential~\eqref{VH} are 
\begin{align}
&M_{A}^2=\frac{m_{12}^2}{s_\beta c_\beta}~,\qquad\qquad
M_{H^\pm}^2=M_{A}^2-\frac{1}{2}\lambda_{12} v^2,\notag\\
&M_{h,H}^2=\frac{1}{2}\left(M_{A}^2\pm\sqrt{M_{A}^4-8M_{H^\pm}^2\lambda_{12} v^2s_\beta^2 c_\beta^2}~\right),
\end{align}
which results in a spectrum $M_{A}>M_{H}>M_{H^\pm}$. This is different from the 2HDM spectrum of the MSSM because the dominant quartic term is $\lambda_{12}$. For the benchmark point of the previous section, the three masses are estimated to be
\begin{align}
M_{A} \sim 380\text{ GeV},\quad
M_{H} \sim 370\text{ GeV},\quad\text{and}\quad
M_{H^\pm} \sim 340\text{ GeV}.
\end{align}

\subsection{Other pNGBs}
In addition to the two doublets, the pNGBs also include a real triplet $\phi$, a real singlet $\eta$, and a complex singlet $s$.
The triplet obtains its mass from the gauge loop as shown in Eq.~\eqref{mphi}. For $M_\rho\sim 5$ TeV, it gives
\begin{align}
&M_\phi^2=c_w\frac{1}{4\pi^2}g^2M_\rho^2\sim\left(500\text{ GeV}\right)^2.
\end{align}

The singlets do not receive mass contributions from SM gauge interactions. The complex singlet $s$ obtains its mass from 
 the collective symmetry breaking mechanism \eqref{Vcol},
\begin{equation}\label{smass}
M_s^2=c_{k\ell}\frac{N_c}{4\pi^2}\lambda_{k_L}^2\lambda_{\ell_L}'^2f^2+
c_{mn}\frac{N'_c}{4\pi^2}\lambda_{m_L}^2\lambda_{n_L}'^2f^2\geq 4\lambda_{12}f^2\approx (2f)^2,
\end{equation}
which is expected to be at the TeV scale. There is also a tadpole term from the PQ-violating potential, which will introduce a small  VEV for $s$,
\begin{equation}\label{svev}
\langle s\rangle\sim \frac{m_{12}^2f}{M_s^2}\leq \frac{\left(210\text{ GeV}\right)^2}{4f}\sim \mathcal{O}(10 \text{ GeV}).
\end{equation}
It will have little effect on the mass of the singlet.

Finally, the real singlet $\eta$ does not get a mass at the leading order but it couples quadratically to the Higgs doublets (e.g., from Eq.~\eqref{H2_mass}), so it can still become massive after the Higgs doublets develop nonzero VEVs. Through Eq.~\eqref{H2_mass},  $\eta$ receives a mass
\begin{align}
M_\eta^2\sim \frac{3}{8\pi^2}y_t^2M_T^2\cdot \left(\frac{v}{f}\right)^2\implies
M_\eta \sim \left(\frac{M_T}{f}\right) 48\text{ GeV}.
\end{align}
For naturalness, a relatively light top partner is preferred. On the other hand, the experimental constraints require $\eta$ to be heavier than half of Higgs boson mass to avoid large Higgs decay rate to the $\eta\eta$ channel. We expect a light singlet scalar around 100 GeV, which can be the lightest composite state in the spectrum.

\section{Collider Searches}\label{sec:Collider}

In CHMs, there will be new composite states of scalars, fermions, and vectors near or below the compositeness scale. The detailed spectrum and quantum numbers depend on the specific realizations of the CHMs. In this section, we study the collider searches of and constraints on these new states in the $SU(6)/Sp(6)$ model discussed in this paper.

\subsection{The Second Higgs Doublet}

\begin{figure}[tbp]
\centering
\includegraphics[width=0.6\textwidth]{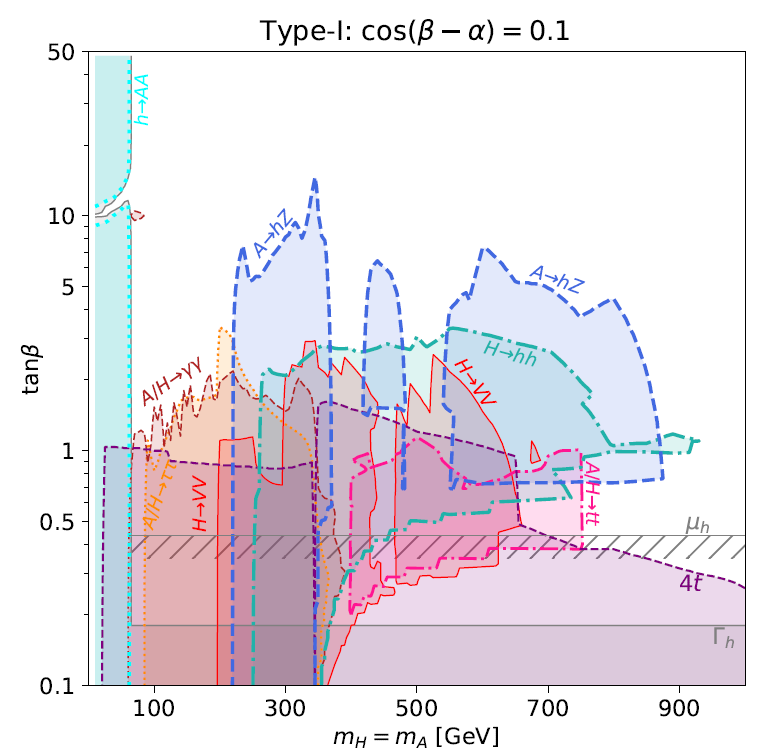}
\caption{\label{Hbound} Constraints on extra neutral Higgs bosons in a Type-I 2HDM with a small misalignment $c_{\beta-\alpha}=0.1$. This summary plot is taken from Ref.~\cite{Kling:2020hmi}.}
\end{figure}

Under the requirement of naturalness, the second Higgs doublet is expected to be among the lightest states of the new resonances and could be the first sign of this model. In the Type-II 2HDM, the flavor-changing process $b\to s\gamma$ has put strong constraints on the charged Higgs mass to be above 600 GeV, which would require more tuning in the Higgs potential. Therefore, we focus on the Type-I 2HDM scenario. As explained in the previous section, we will consider a relatively small $\tan \beta \sim 3$ with a small misalignment $c_{\beta-\alpha}\sim 0.1$.

The direct searches can be divided into two categories -- charged Higgs bosons $H^\pm$ and neutral Higgs bosons $H^0, A^0$. In the Type-I 2HDM with a small misalignment, neutral Higgs bosons to fermion couplings are characterized by a factor $-s_\alpha/s_\beta\sim 1/4$ and the charged Higgs boson to fermion couplings are characterized by $c_\beta/s_\beta\sim 1/3$. Comparing to neutral Higgs bosons, the charged Higgs boson searches give a more reliable constraint on tan$\beta$ because it doesn't depend on the mixing angle $\alpha$.

The charged Higgs boson is searched by its decays to SM fermions. For $M_{H^\pm} \lesssim m_t$, the strongest constraint comes from decaying to $\tau \nu$~\cite{Aaboud:2018gjj,Sirunyan:2019hkq}. Interpreted in the Type-I model, it excludes $\tan\beta <14$ for $M_{H^\pm}\sim 100$~GeV and $\tan\beta <3$ for $M_{H^\pm}$ up to 150~GeV~\cite{Sanyal:2019xcp}. For a heavier charged Higgs boson,  the main constraint comes from the decay to $tb$, which rules out $\tan\beta \lesssim 2$ for $M_{H^\pm}$ in the range of 200-400 GeV, and becomes weaker for larger $M_{H^\pm}$~\cite{Aaboud:2018cwk, Sirunyan:2019arl}.

For neutral Higgs bosons, there are multiple decay channels being searched. For light states below the $t\bar{t}$ threshold, they can be searched by $H/A\to \tau\tau$~\cite{Sirunyan:2018zut, Aad:2020zxo} and $H\to \gamma\gamma$~\cite{Sirunyan:2018wnk, Aaboud:2017yyg} decays. For heavier states, the decay to $t\bar{t}$ becomes accessible and dominant. The searches of $H/A\to t\bar{t}$ has been done at CMS and ATLAS \cite{Sirunyan:2019wph,Aaboud:2018mjh}. These searches typically constrain $\tan\beta \gtrsim 1-2$ up to $M_{H/A}\sim 750$~GeV. When there is misalignment as expected in this model, there are also additional decay channels of these neutral scalars which give important constraints. These include $H/A\to WW$~\cite{Sirunyan:2019pqw,Aaboud:2017gsl} and $ZZ$~\cite{Sirunyan:2018qlb,Aaboud:2017rel}, $H\to hh$~ \cite{Sirunyan:2018ayu, Aad:2019uzh}, and $A\to hZ$~ \cite{Sirunyan:2019xls, Aaboud:2017cxo}. The $A\to hZ$ and $H\to hh$ turn out to be most constraining for the region that we are interested in. The $A\to hZ$ can exclude $\tan\beta$ up to 10 below the $t\bar{t}$ threshold. Some higher mass ranges are also constrained due to data fluctuations. $H\to hh$ constrains $\tan\beta$ to be $\gtrsim 3$ for a wide mass range. Various constraints on the neutral scalars for 2HDMs are summarized in Ref.~\cite{Kling:2020hmi}, and the relevant plot is reproduced in Fig.~\ref{Hbound}. We can see that the benchmark point chosen in the previous section,
\begin{align}
M_{A} \sim 380\text{ GeV},\quad
M_{H} \sim 370\text{ GeV},\quad\text{and}\quad
M_{H^\pm} \sim 340\text{ GeV},
\end{align}
with $\tan\beta=3$ is sitting in the gap of the constraints. It is still allowed by but very close to the current constraints, hence it will be tested in the near future.

For future searches, the most relevant channels for the more natural mass range are di-boson channels $H/A\to VV$, $H\to hh$, and $A\to hZ$. The current bounds are expected to be improved by $\sim 10$ times~\cite{Cepeda:2019klc}. It will probe the parameter region that we are most interested in. If we can also find the charged Higgs with a slightly lighter mass, this particular spectrum can be an indication of the specific 2HDM Higgs potential (different from that of the MSSM) that arises from this type of CHMs.

\subsection{Additional Scalar Bosons}

Besides the second Higgs doublet, there are also several additional scalar bosons, which include a real triplet $\phi$, a complex singlet $s$, and a real singlet $\eta$. At the leading order, they don't directly connect to the SM fermions. However, the couplings to SM fermions are induced through the mixing with Higgs bosons after EWSB, with a suppression factor of $v/2f \sim 0.15$ (for $\xi\sim 0.1$).

\paragraph{Scalar triplet $\phi$:}

The scalar triplet has unsuppressed gauge interactions with $W$ and $Z$ bosons, but only through four-point vertices. They can be paired produced through the vector boson fusion but the production is highly suppressed due to the large energy required. Therefore, here we only consider the single production through the interaction with SM fermions. The scalar triplet includes a complex charged scalar $\phi^{\pm}$ and a neutral scalar $\phi^0$. The collider searches of the charged scalar are similar to those of $H^{\pm}$ of the second Higgs doublet but with the suppressed couplings. It can be produced in association with a top and a bottom. However, due to the suppressed coupling and the larger mass, the charged scalar $\phi^{\pm}$ is less constrained.

The neutral scalar $\phi^0$ is searched in the same ways as the neutral scalars in 2HDMs. Guided by the benchmark scenario, we consider a scalar with mass $\sim 500$ GeV, which gives a cross section $120$ fb. The dominant decay mode will be
$\phi^0\to t \bar{t}$ with a branching ratio $\sim 75\%$. The current bound from the LHC searches~\cite{Sirunyan:2019wph,Aaboud:2018mjh} on the cross section is $\sigma\times BR<5$ pb, which is still loose for a neutral scalar with $\sigma\times BR\sim 90$ fb. The di-boson modes are also important with branching ratios $\sim 16\%$ for $WW$ and $\sim 8\%$ for $ZZ$. The most stringent current upper bound comes from $\phi^0\to ZZ$ channel, which ruled out $\sigma \times BR$ above $100$ fb \cite{Sirunyan:2018qlb,Aaboud:2017rel}. It is also much larger than $\sim 10$ fb for the benchmark point. In the future, around $3.6\times10^5$ $\phi^0$ (at 500 GeV) would be produced in the HL-LHC era with an integrated luminosity of 3 ab$^{-1}$. The bound can be improved by $10$ times~\cite{Cepeda:2019klc}. And a 500 GeV $\phi^0$ could be within reach in the HL-LHC era.

\paragraph{Scalar singlets:}

The complex scalar $s$ is expected to be at TeV scale and the real singlet $\eta$ is around $100$ GeV. They both act like the neutral scalar $\phi^0$ discussed above, but without the gauge interactions. They can be produced through the gluon fusion but the production cross sections will be suppressed by $\xi/4 \sim 0.025$.

For the heavy complex scalar $s$, The expectation of its mass in the benchmark point is above $1.5$ TeV. The dominant decay channel will be a pair of neutral Higgs bosons $s\to h_1h_2~(hh,hH,HH)$ or  charged Higgs bosons due to the large $s\tilde{H_2}^\dagger H_1$ coupling. It also connects to the fermions sector through the mixing with Higgs bosons. However, the production is suppressed due to the large mass. Although it is an essential element of the collective Higgs quartic term, it is hard to detect even at the HL-LHC. It may be accessible in the next generation hadron collider.

The light real scalar $\eta$ should be heavy enough so that $h\to\eta\eta$ is forbidden due to the constraint from the Higgs invisible decay measurement~\cite{Khachatryan:2016whc}. This requires $M_T/f \gtrsim 1.3$ for a realistic model, but it should remain relatively light if the top partner is not too heavy for the naturalness reason. Since the interactions between $\eta$ and SM particles are all through the mixing with the Higgs boson, the search modes are similar but with the $\xi/4$ suppression on the production rate. The cross section is $\sim1.5$ pb for a 100 GeV $\eta$.  The dominant decay modes are $b\bar{b}\, (78.9\%)$, $\tau\tau\,(8.3\%)$ and $gg\,(7.4\%)$, but they all suffer from large backgrounds. On the other hand, the clean channel $\gamma\gamma$ suffers from a low branching ratio $\sim 0.16\%$. For the benchmark point, the diphoton channel has $\sigma\times BR\sim 3$ fb. The latest  search from CMS \cite{CMS:2017yta} still has an uncertainty $\sim 20 (10)$ fb for a diphoton invariant mass $\sim 80 (110)$ GeV, much bigger than the cross section that we expect. With more data and improvements in the background determinations, it might be discoverable at future LHC runs.

\subsection{Fermionic Top Partners}
The top partners in the $SU(6)/Sp(6)$ CHM are vector-like fermionic resonances which form a sextet of the $Sp(6)$ global symmetry. Their quantum numbers under the SM gauge symmetry are $(3,2,1/6) [\times 2]$, $(3,1,2/3)$, and $(3,1,-1/3)$, which are identical to those of SM quarks. There are no exotic states with higher or lower hypercharges. These states are degenerate in the limit of unbroken $Sp(6)$ global symmetry. (Small splittings arise from the explicit symmetry breaking effects and EWSB.) Their mass $M_T$ plays the important role of cutting off the quadratic contribution from the top quark loop to the Higgs potential. Naturalness prefers $M_T$ to be as low as possible allowed by the experimental constraints. The current bound on the top partner mass has reached $\sim 1.2$~TeV~\cite{Sirunyan:2018omb, Aaboud:2018pii}. 
The HL-LHC can further constrain the mass up to $\sim 1.5$~TeV~\cite{CidVidal:2018eel}. The benchmark value of 1.6 TeV is close to but probably still beyond the reach of HL-LHC.  A future 100 TeV collider will cover the entire interesting mass range of the top partners if no severe tuning conspires. It may even be able to find the fermionic partners of the other SM quarks, which are expected to be much heavier.

\subsection{Heavy Vector Bosons}
Unlike the top partners, the partners of SM gauge bosons (spin-1 resonances) are not necessarily light because of the smallness of $SU(2)_W$, $U(1)_Y$ gauge couplings. In fact, their masses need to be large enough to give a sufficiently large mass to the second Higgs doublet and to cancel in a large part the negative contribution from the top sector to the quadratic Higgs potential. The largest couplings of these composite spin-1 resonances are to the composite states, including the pNGBs. Their mixings with SM gauge bosons are strongly suppressed by their multi-TeV masses, hence their couplings to SM light fermions are also suppressed, resulting in a small production rate as well as small decay branching ratios to SM elementary particles~\cite{Pappadopulo:2014qza,Greco:2014aza}. The leading decay modes will be through the composite states, such as top partners or pNGBs which include the longitudinal modes of $W$ and $Z$. The current searches of heavy vector triplets decaying into SM gauge bosons final states have reached a bound about 4 TeV~\cite{Aad:2020ddw, Aad:2019fbh,Sirunyan:2019vgt,Sirunyan:2019jbg}. The bound is  relieved for larger $g_\rho>3$ with more suppression on the production rate. Besides, the model contains a richer sector of the pNGBs which will dilute the decay branching fractions to SM gauge bosons, further reducing the bound. If the vector resonances are heavier than twice the top partner mass, the decaying into top partners will dominate and it would require different search strategies. As the production rate quickly diminishes for heavier vector resonances, the typically expected masses of the vector resonances as in our benchmark will be out of reach even at the HL-LHC. A future higher energy machine will be needed to discover them.

\section{Precision Tests}\label{sec:Precision}
In this section, we discuss the indirect tests of this model from precision experimental measurements.

\subsection{Higgs Coupling Measurements}

The Higgs boson couplings to SM fields in the $SU(6)/Sp(6)$ CHM are modified by two effects: the nonlinear effect due to the pNGB nature of the Higgs boson and the misalignment from the mixing of the 2HDM.
The deviation of the Higgs coupling to vector bosons is parameterized by
\begin{equation}
\kappa_V\equiv \frac{g_{hVV}}{g^{SM}_{hVV}}=
\text{sin}(\beta-\alpha)~\text{cos}\frac{\sqrt{V_1^2+V_2^2}}{\sqrt{2}f}~,
\end{equation}
where the first factor comes from the misalignment of the 2HDM and the second factor is the nonlinear effect of the pNGB. For the benchmark point in Sec.~\ref{sec:2HDM}, $\sin(\beta -\alpha) \approx 0.995$, which gives  
\begin{equation}
\kappa_V
\approx (0.995)\sqrt{1-\frac{\xi}{2}}\approx 0.995-0.249~\xi~,
\end{equation}

The deviation of the Higgs coupling to fermion is universal in Type-I 2HDMs because it couples to all fermions in the same way. The expression is somewhat more complicated in CHM, and here we only expand to $\mathcal{O}(\xi)$,
\begin{equation}
\kappa_f\equiv \frac{g_{hff}}{g^{SM}_{hff}}=
\frac{1}{s_\beta}\left(c_\alpha-\xi~\frac{1}{12}(3s_\beta^2c_\alpha+c_\beta^2c_\alpha-2s_\beta c_\beta s_\alpha)\right)
\approx 1.030-0.252~\xi~,
\end{equation}
where the numerical value of the last expression is obtained for the benchmark point.

\begin{figure}[tbp]
\centering
\includegraphics[width=.7\textwidth]{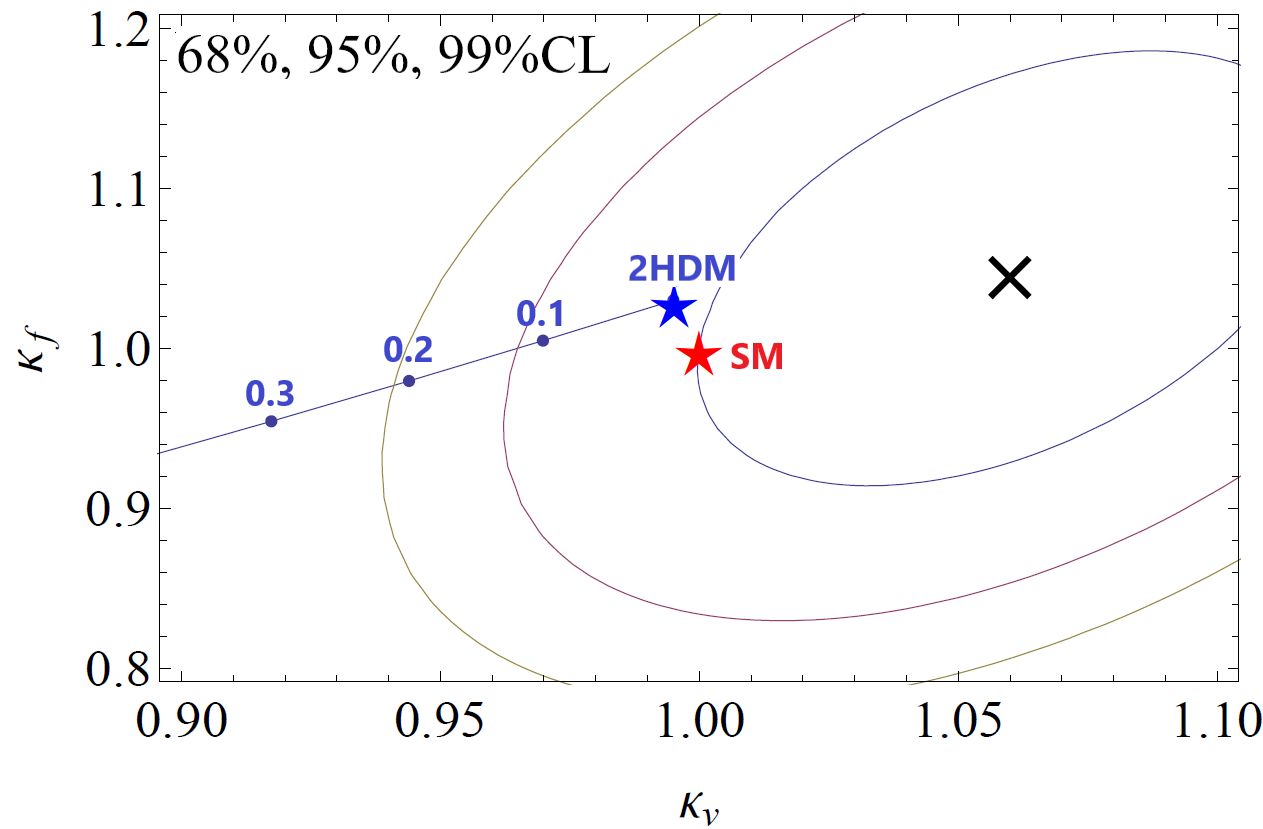}
\caption{\label{HiggsCoupling} The fit of the Higgs coupling strengths to the gauge bosons ($\kappa_V$) and fermions ($\kappa_f$) obtained by the ATLAS \cite{Aad:2019mbh} from the 13 TeV LHC data. The cross is the observed central value. The circles from inside out represent the 68\%, 95\%, and 99\% CL respectively. The red star shows the SM value $(1,1)$. The blue star in the predicted value of the 2HDM benchmark of Sec.~\ref{sec:2HDM} with $\xi=0$. Along the line, we show the predictions for the same benchmark with different $\xi$ from 0 to 0.3.}
\end{figure}

The current best-fit values of $\kappa_V$ and $\kappa_F$ from ATLAS \cite{Aad:2019mbh} with an integrated luminosity of 80 fb$^{-1}$ are
\begin{align}
&\kappa_V=1.06\pm 0.04~,\\
&\kappa_F=1.05\pm 0.09~,
\end{align}
with a 45\% correlation between the two quantities. The central values for both quantities are slightly above the SM value 1, but without significant deviations given the uncertainties.  As shown in Fig.~\ref{HiggsCoupling}, within 95\% CL level, $\xi\leq 0.12$ is still allowed (for the benchmark point), which gives a lower bound on the scale $f\sim$ 700 GeV.

In the future,  the uncertainties in $\kappa_V$ and $\kappa_F$ can be improved to 1\% and 3\% respectively at the HL-LHC,~\cite{deBlas:2019rxi}. Assuming the central values of $(1,1)$, it can bound $\xi$ down to $0.1$ at 99\% CL. The next generation Higgs factories, such as ILC, CEPC, and FCCee, will have great sensitivities to the $hZZ$ coupling and can measure $\kappa_V$ with a precision $\approx$ 0.3\%. It can test the scale $f$ up to several TeV and hence cover the entire natural parameter region for the CHMs.

Another decay mode worth mentioning is $h\to \gamma\gamma$. The branching ratio of this decay mode will receive an additional contribution from charge Higgs bosons. But the current bound from this decay mode is still loose. It will improve at HL-LHC and  future Higgs factories. It may provide a sign of the heavy charged Higgs bosons if they exist.

\subsection{Flavor Changing Neutral Currents}\label{Flavor}
New physics appearing near the TeV scale may introduce dangerously large flavor changing neutral currents (FCNCs), so the flavor-changing processes put strong constraints on the model constructions. The $SU(6)/Sp(6)$ model contains two light Higgs doublets. If general Yukawa couplings are allowed between them and SM fermions, large FCNCs will be induced. Therefore, it is desirable to impose the natural flavor conservation such that each type of Yukawa couplings only comes from one of the two Higgs doublets. Even so, a light charged Higgs boson can induce a significant contribution to the branching ratio BR$(B\to X_s\gamma)$~\cite{Grinstein:1987pu,Hou:1987kf,Rizzo:1987km,Geng:1988bq,Barger:1989fj,Grinstein:1990tj}. In the Type-II or flipped 2HDM, this gives a lower bound on the charged Higgs boson $M_{H^\pm}>600$ GeV~\cite{Arbey:2017gmh,Misiak:2017bgg}, which would introduce more tuning in the Higgs potential. To have a more natural model, we therefore focus on the construction of the Type-I 2HDM. In a Type-I model, the $B\to X_s\gamma$ constraint rule out the region below $\tan\beta <2$~\cite{Arbey:2017gmh,Misiak:2017bgg}.

The partial compositeness couplings between the elementary fermions and the composite operators can potentially induce FCNCs. In our construction, the largest such couplings (for the top Yukawa and the collective Higgs quartic term) preserve a Peccei-Quinn symmetry with different PQ charges for different generations (see Appendix \ref{PQ charge}). As a result, there is no FCNC induced by these large couplings in the leading order. Some FCNCs may be induced by other (smaller) couplings which are responsible for generating the complete SM fermion masses and mixings, but they are suppressed by the small couplings and depend on the details of their pattern.

\subsection{Oblique Parameters}

The electroweak oblique corrections provide important tests of new physics near the weak scale. They are usually expressed in terms of $S$, $T$, and $U$ parameters~\cite{Peskin:1990zt,Peskin:1991sw}. The current global fit gives~\cite{Tanabashi:2018oca}
\begin{equation}
S=-0.01 \pm 0.10, \quad T=0.03 \pm 0.12, \quad U= 0.02\pm 0.11.
\end{equation}
For heavy new physics, $U$ is typically small as it  is suppressed by an additional factor $M_{\rm new}^2/m_Z^2$. If one fixes $U=0$, then $S$ and $T$ constraints improve to
\begin{equation}
S=0.0 \pm 0.07, \quad T=0.05 \pm 0.06,
\end{equation}
with a strong positive correlation (92\%) between them. At 95\% CL, one obtains $S<0.14$ and $T<0.22$.

There are several contributions to the oblique parameters in our model, with similarities and differences compared to the MCHM discussed in the literature. First, our model has two Higgs doublets. Their contributions to $S$ and $T$ can be found in Ref.~\cite{Haber:1993wf,Haber:1999zh,Froggatt:1991qw}. To satisfy the other experimental constraints, the Higgs potential needs to be close to the alignment limit and the heavy states are approximately degenerate. The contributions are expected to be small and do not provide a significant constraint~\cite{Haller:2018nnx}. The other contributions are discussed below.

\subsubsection*{The $S$ parameter} 
The leading contribution to the $S$ parameter comes from the mixing between the SM gauge bosons and the composite vector resonances. It is estimated to be ~\cite{Agashe:2003zs,Agashe:2005dk,Giudice:2007fh}
\begin{align}
\Delta S \sim c_S~4\pi \frac{v^2}{M_\rho^2}  
\sim c_S~0.03 \left(\frac{5 \text{ TeV}}{M_\rho}\right)^{2},
\end{align}
where $c_S$ is an $\mathcal{O}(1)$ factor. It gives a lower bound of $\sim 2.5$~TeV on $M_\rho$ for $c_S=1$.

In CHMs, there is a contribution from the nonlinear Higgs dynamics due to the deviations of the Higgs couplings, which result in an incomplete cancellation of the electroweak loops~\cite{Barbieri:2007bh,Grojean:2013qca}.  This contribution is proportional to $\xi$ and depends logarithmically on $M_\rho/M_h$. For $M_\rho=5$~TeV, it gives $\Delta S\sim 0.10 ~\xi$ which is well within the uncertainty.\footnote{A factor of 1/2 is included due to the normalization of $f$ compared to Ref.~\cite{Barbieri:2007bh,Grojean:2013qca}.} In the MCHM, there is also a contribution due to loops of light fermionic resonances. It is logarithmically divergent and its coefficient depends on the UV physics~\cite{Grojean:2013qca}. This contribution can be significant, depending on the UV-sensitive coefficient. However, in our model, the fermionic resonances are complete multiplets of $SU(6)$ and their kinetic terms remain $SU(6)$ symmetric, so this divergent contribution is absent.

\subsubsection*{The $T$ parameter} 
The $T$ parameter parametrizes the amount of custodial $SU(2)$ breaking. There are also several potential contributions in our model. First, the pNGB spectrum contains a real $SU(2)_W$ triplet $\phi$. If it obtains a VEV induced by the trilinear scalar couplings to a pair of Higgs doublets, $H_1^\dagger \phi H_1$, $H_2^\dagger \phi H_2$, or ($H_1 \phi H_2+\text{h.c.}$), it will give a tree-level contribution to $\Delta T$. Its VEV is bounded to be less than $\sim 8$~GeV, putting strong constraints on these couplings. However, if all the large couplings are real and the CP symmetry is (approximately) preserved, the real scalars $\phi$ and $\eta$ are CP odd and the interactions $H_1^\dagger \phi H_1$, $H_2^\dagger \phi H_2$, and ($H_1 \phi H_2+\text{h.c.})$ are forbidden by the CP symmetry. The $\eta$ and $\phi$ fields need to couple quadratically to the Higgs fields. This also justifies the Higgs potential analysis based on the 2HDM potential. Of course, CP symmetry has to be broken in order to allow the nonzero phase in the CKM matrix. We assume that this is achieved with the small partial compositeness couplings so that the induced trilinear scalar couplings are kept small enough to satisfy the bound.

Apart from the potential triplet VEV contribution, the leading contribution to $\Delta T$ comes from fermion loops. For the partial compositeness couplings in this model, the custodial symmetry breaking comes from $\lambda_R$.\footnote{The custodial symmetry of our model corresponds to the Case B in Ref.~\cite{Cai:2018tet}} The dominant contribution comes from the light top partners and the corresponding mixing coupling $\lambda_{t_R}$ The deviation is estimated to be~\cite{Giudice:2007fh}\footnote{The partial compositeness couplings are related to the top Yukawa coupling by $\lambda_{t_L}\lambda_{t_R}\sim y_t \,g_T$. For $y_t s_\beta \sim 0.85$ at 2 TeV and assuming $g_T~\sim 2$, we need  $\sqrt{\lambda_{t_L}\lambda_{t_R}}\sim 1.3$.}
\begin{align}
\Delta T \sim \frac{N_c}{16\pi^2\alpha} \lambda_{t_R}^4 \frac{v^2}{M_T^2}
\sim 0.16 \left(\frac{\lambda_{t_R}}{1.3}\right)^4 \left(\frac{1.6 \text{ TeV}}{M_T}\right)^2.
\end{align}

There is also a contribution from the modifications of the Higgs couplings to gauge bosons due to the nonlinear effects of the pNGB Higgs. The contribution to $\Delta T$ from the nonlinear effects again depends on $\xi$ and is logarithmically sensitive to $M_\rho$. For $M_\rho=5$~TeV, it gives $\Delta T \sim -0.28 ~\xi$~\cite{Barbieri:2007bh,Grojean:2013qca}. It is significant and can partially cancel the light top partner contribution. The contribution from the mixing of the hypercharge gauge boson and vector resonances is small due to the custodial symmetry. The tree-level contribution vanishes and the loop contribution is negligible. The overall $\Delta T$ correction is expected to be positive and could help to improve the electroweak precision fit in the presence of a positive $\Delta S$.

In summary, among the various sources of the corrections to the electroweak observables, the contributions from the composite resonances are expected to be dominant. They give strong constraints on the masses of heavy resonances $M_\rho$ and $M_T$ as well as the relevant coupling like $\lambda_{t_R}$. Nevertheless, for natural parameter values as our benchmark, the corrections on $(S,T)$ can still lie safely within the current uncertainty region. A future $Z$ factory can greatly improve the precisions of the electroweak observables, which can provide a strong test of the model.

\subsection{$Zf\bar{f}$ Couplings}\label{Zff}

The partial compositeness couplings generate mixings between elementary fermions and composite resonances. They can modify the $Zf\bar{f}$ couplings in the SM. This is a well-known problem in CHMs for the $Zb\bar{b}$ coupling in implementing the top partial compositeness. A solution based on an extended custodial symmetry $SU(2)_V\times P_{LR}$ on the top sector by embedding the left-handed top-bottom doublet into the $(2,2)$ representation of $SU(2)_L\times SU(2)_R$ was proposed in Ref.~\cite{Agashe:2006at}. The top sector in our construction does not have this extended custodial symmetry. Furthermore, to obtain the collective quartic Higgs term, we need several large partial compositeness couplings involving other light SM fermions. As a consequence, we may expect significant deviations of the $Zf\bar{f}$ couplings for all fermions involved and they present important constraints on this model.

The third generation left-handed quark's partial compositeness couplings modify the $Zb_L\bar{b}_L$ coupling. Its deviation $\delta g_{b_L}$ from the current experimental determination is constrained within $3\times 10^{-3}$~\cite{Batell:2012ca}. This deviation comes from mixings between the bottom quark $b$ and the corresponding composite resonances $B$. Under our assignment in Appendix \ref{PQ charge}, there are two terms that will have large positive contributions to $\delta g_{b_L}$. They are
\begin{align}
\lambda_{t_L}\bar{q}_{3,L}H_1B_{R} \quad\to\quad (\lambda_{t_L}v_1)\bar{b}_LB_{R}~,\\
\lambda_{b_L}'\bar{q}_{3,L}\tilde{H_2}B'_{R} \quad\to\quad (\lambda_{b_L}'v_2)\bar{b}_LB'_{R}~.
\end{align}
The first one is responsible for generating the top Yukawa coupling and induces the mixing between $b_L$ and the bottom partner $B$ with PQ charge 0. The second introduces the bottom Yukawa coupling and the collective quartic term. It induces the mixing with another bottom partner $B'$ with PQ charge 1. The deviations that they bring can be estimated as 
\begin{align}
\delta g_{b_L}\approx \frac{\lambda_{t_L}^2c_\beta^2}{M_0^2(\text{TeV})}\times (30\times 10^{-3}),\quad
\delta g_{b_L}\approx \frac{\lambda_{b_L}'^2s_\beta^2}{M_1^2(\text{TeV})}\times (30\times 10^{-3})~,
\end{align}
where $M_0$ and $M_1$ are the masses of the fermions resonances $B$ and $B'$ respectively. Note that $M_0$ is also the top partner mass which is responsible to cut off the top loop contribution to the quadratic Higgs potential so it should not be too large for naturalness. On the other hand $M_1$ is the bottom partner mass which can be much larger because of the small bottom Yukawa coupling.
These corrections impose strong constraints on the couplings and masses of the composite fermion resonances. For the first term, taking $\lambda_{t_L}\approx 1.3$ and $c^2_\beta\approx 0.1$ from the benchmark model, it requires $M_0=M_T\gtrsim 1.3$ TeV, which is still in the range we expect. Compared to the other models without the $SU(2)_V\times P_{LR}$ custodial symmetry, such as the MCHM$_4$~\cite{Agashe:2004rs}, we are saved by the $c_\beta^2$ factor to allow a relatively light top partner.
For the second one, taking $\lambda'_{b_L}\approx 1.9$ and $s_\beta\approx 1$ would require $M_1\gtrsim 6$ TeV for the bottom partner. The bound on $M_1$ can be reduced for a smaller value of $\lambda'_{b_L}$, but at the cost of a larger $\lambda_{c_L}$ if their combination is responsible for the collective Higgs quartic term, which increases the deviations for $\delta g_{c_L}$ and $\delta g_{s_L}$.

The collective Higgs quartic term needs at least four large $\lambda_{L}, \lambda'_{L}$ couplings. Each of them will induce two $\delta g_{L}$ deviations from SM $Zf\bar{f}$ couplings and all of them reduce the magnitudes from the SM predicted values. Since the $Z$ decay width and branching ratios are all well measured at $\mathcal{O}(10^{-3})$ precision, we also need to examine their observable consequences and the corresponding constraints.

It is harder to extract the constraints on individual couplings from the observables that depend on more complicated combinations of different couplings. Therefore we consider the constraints from $\Gamma$(hadron) and $\Gamma$(charged lepton) because they are directly proportional to the couplings instead of some ratios. We predict  smaller values for both $\Gamma$(hadron) and $\Gamma$(charged lepton), but their observed central values are both larger than the SM predictions so the allowed parameter space is strongly restricted. At the 95\% CL level, the allowed negative deviations are \cite{Tanabashi:2018oca}
\begin{align}
\Delta\Gamma(\text{had})\sim -1.0 \text{ MeV ,}\qquad
\Delta\Gamma(\ell^+\ell^-)\sim -0.15 \text{ MeV .}
\end{align}
From these, we obtain the constraints on allowed negative deviations on the magnitude of different left-handed fermion couplings (assuming only one term dominates) as follow,
\begin{subequations}
\begin{align}
|\delta g_{u_L}|<0.7 \times 10^{-3} &\quad\text{for up-type quarks,}\\
|\delta g_{d_L}|<0.6 \times 10^{-3} &\quad\text{for down-type quarks,}\\
|\delta g_{e_L}|<0.4 \times 10^{-3} &\quad\text{for charged leptons.}
\end{align}
They strongly constrain the parameters of our model. To satisfy these constraints, the corresponding fermion partners need to be over 10 TeV if their couplings to the elementary fermions are large enough to be responsible for the collective Higgs quartic term. 

These constraints can be relaxed somewhat if we use the neutrino couplings for the collective Higgs quartic term. The $\Gamma$(invisible) is smaller than the SM prediction. The allowed negative deviation is 4 MeV at the 95\% CL level, which corresponds to
\begin{align}
|\delta g_{\nu_L}|<6 \times 10^{-3} &\quad\text{for neutrinos.}
\end{align}
\end{subequations}
The resulting constraints on the corresponding fermion resonances are milder.

The precision measurements of the $Z$ couplings put strong constraints on our model because we predict a reduction of all $Z f_L \bar{f_L}$ couplings in the construction. A future $Z$ factory may improve the coupling measurements by more than one order of magnitude. Consequently, it can either establish a deviation from the SM predictions which points to new physics in the nearby scales, or further affirm the SM predictions which will severely challenge this model or any other models with similar predictions. Nevertheless, we would like to emphasize that these constraints are indirect so it is quite possible that one can extend the model to introduce new contributions to cancel the deviations, at the expense of complexity and/or tuning.

\section{Conclusions}\label{sec:Conclusion}

Composite Higgs models remain an appealing solution to the hierarchy problem. However, in realistic models, some tuning in the Higgs potential is often required to obtain the correct EWSB and the observed Higgs boson mass. One source is from the mass splittings within the top partner multiplet of the composite resonances, which can generate a large quadratic Higgs potential through the partial compositeness couplings at the order $\lambda_{L(R)}^2$. The other is to obtain the necessary relative size between the quartic term and the quadratic term of the Higgs potential in order to separate the EWSB scale and the compositeness scale. In this paper, we look for models that can address both problems. We show that a CHM based on the coset $SU(6)/Sp(6)$ can achieve the goals without introducing additional elementary fields beyond the SM and the composite sector, which otherwise will introduce a new coincidence problem that why the new elementary fields and the compositeness resonances are at the same mass scale.

A key part of the setup is to couple the elementary SM fermions to the composite operators of the fundamental representation of $SU(6)$. The composite resonances do not split after the symmetry is broken to $Sp(6)$ and hence do not induce any large potential from the UV dynamics for the pNGBs. The leading contribution to the Higgs quadratic term is reduced to the unavoidable top quark loop in the IR. In addition, the fundamental representation of $SU(6)$ contains two electroweak doublets of the same SM quantum numbers. This allows us to write down different ways of coupling between the elementary fermions and the composite resonances, each of which preserves a subset of the global symmetry. In this way, a quartic Higgs potential can be generated from the collective symmetry breaking of the little Higgs mechanism, without inducing the corresponding quadratic terms. This independent quartic term enables us to naturally separate the EWSB scale and the $SU(6)$ global symmetry breaking scale, reducing the tuning of the Higgs potential.

This model contains many more pNGBs than one Higgs double of the minimal model. In particular, there are two Higgs doublets and the second Higgs doublet should not be too heavy for naturalness considerations. The extra Higgs bosons are already subject to collider constraints and are the most likely new particles to be probed in the future LHC runs beside the top partners. The other pNGBs, having smaller couplings to SM particles, are more difficult to find. Together with the heavy vector and fermion resonances, they need higher energy machines with large integrated luminosities. The top partners in this model do not include new particles with exotic charges, e.g., 5/3, as in many other CHMs. The model also predicts deviations of the Higgs couplings and weak gauge boson couplings. The current experimental data already provide substantial constraints on the model parameters in the most natural region. The Higgs coupling measurements will be greatly improved at the HL-LHC and future Higgs factories. A future $Z$ factory can also further constrain the electroweak observables. Either the agreements with SM predictions with higher precisions will push the model completely out of the natural scale for the solution to the hierarchy problem, or some deviations will be discovered to point to the possible new physics, and if any of the CHMs can provide an explanation for them.

\acknowledgments

We thank Felix Kling, Shufang Su, and Wei Su for letting us use the figure in Ref.~\cite{Kling:2020hmi}. We also thank Da Liu, Matthew Low, and Ennio Salvioni for useful discussions. This work is supported by the Department of Energy Grant number DE-SC-0009999.

\appendix
\section{The $SU(5)/SO(5)$ Composite Higgs Model}\label{SU(5)/SO(5)}

The $SU(5)/SO(5)$ is also a possible coset that can naturally avoid large UV contributions to the Higgs potential. It was one of the cosets considered in early composite Higgs models of 1980s~\cite{Georgi:1984af, Dugan:1984hq}. It was also the coset of the littlest Higgs model~\cite{ArkaniHamed:2002qy} which was one of the pioneer models to realize the mechanism of the collective symmetry breaking for the Higgs quartic coupling. The symmetry breaking can be parametrized by a symmetric tensor field with a VEV
\begin{equation}
\langle \Sigma \rangle = \Sigma_0=
\begin{pmatrix}
0  &  0 & \mathbb{I} \\
0 & 1 & 0 \\
\mathbb{I}   &0 &  0 \\
\end{pmatrix}, \quad \text{where $\mathbb{I}$ is the $2\times 2$ identity matrix.}
\end{equation}
The SM $SU(2)_W$ and $U(1)_Y$ generators are embedded as
\begin{equation}
\frac{1}{2}\begin{pmatrix}
\sigma^a  &  0 & 0 \\
0 & 0 & 0 \\
0  &0 &  - \sigma^{a \ast} \\
\end{pmatrix},
\quad
\frac{1}{2}
\begin{pmatrix}
-\mathbb{I}  &  0 & 0 \\
0 & 0 & 0 \\
0  &0 &  \mathbb{I} \\
\end{pmatrix} + X \mathbf{I}~,
\end{equation}
where the extra $U(1)_X$ charge $X$ accounts for the correct hypercharges of SM fermions.

There are 14 pNGBs, with a complex doublet (which is identified as the Higgs field $H$), a complex triplet $\phi$, a real triplet $\omega$, and a real singlet $\eta$. The partial compositeness couplings can go through the $\mathbf{5}$ and $\mathbf{\bar{5}}$ representations of $SU(5)$. They do not split under $SO(5)$ and hence do not give large UV contributions to the Higgs potential, just as in the $SU(6)/Sp(6)$ case. Under the SM $SU(2)_W \times U(1)_Y$, they decompose as
\begin{subequations}
\begin{equation}
\mathbf{5}_{x}=\mathbf{2}_{x-1/2}\oplus\mathbf{1}_{x}\oplus\mathbf{\bar{2}}_{x+1/2},
\end{equation}
\begin{equation}
\mathbf{\bar{5}}_{x}=\mathbf{\bar{2}}_{x+1/2}\oplus\mathbf{1}_{x}\oplus\mathbf{2}_{x-1/2}~.
\end{equation}
\end{subequations}
To mix with elementary fermions, we need to choose $x=2/3$ for the up-type quarks and $-1/3$ for the down-type quarks.

The Higgs quartic term arising from the collective symmetry breaking takes the form,
\begin{equation} \label{eq:SU(5)collect}
\kappa_1 f^2 \left| \phi_{ij} +\frac{i}{2f}(H_i H_j+H_j H_i)\right|^2 + \kappa_2 f^2 \left| \phi_{ij} -\frac{i}{2f}(H_i H_j+H_j H_i)\right|^2~.
\end{equation}
A drawback of this potential is that a nonzero VEV of the $SU(2)_W$ triplet $\phi$ will be induced after EWSB unless $\kappa_1 = \kappa_2$. The triplet VEV violates the custodial $SU(2)$ symmetry and is subject to the strong constraint of the $T$ (or $\rho$) parameter. Even if we ignore that for a moment, it is also more challenging to generate the collective quartic potential \eqref{eq:SU(5)collect} in this model. The two doublets in $\mathbf{5}$ or $\mathbf{\bar{5}}$ have different hypercharges if $x\neq 0$ and hence are not equivalent. We cannot couple the elementary SM fermion doublets to both $\mathbf{5}$ and $\mathbf{\bar{5}}$ in a way that preserves an $SU(3)$ global symmetry to protect the Higgs mass, so the mechanism introduced for the $SU(6)/Sp(6)$ model in Sec.~\ref{sec:Quartic} does not work here. One could add additional exotic vector-like elementary fermions (with hypercharge $7/6$ or $-5/6$) to couple to these composite operators for the purpose of generating the quartic term, but these exotic elementary fermions should have masses comparable to the compositeness scale, which requires some coincidence. Another possibility is to use the lepton partners that have $x=0$, then the two doublets in $\mathbf{5}$, $\mathbf{\bar{5}}$ are equivalent representations. One can write down the partial compositeness couplings to generate Eq.~\eqref{eq:SU(5)collect}, analogous to the $SU(6)/Sp(6)$ model. However, the same interactions will induce the Majorana mass terms for the left-handed neutrinos through the triplet $\phi$ VEV. The couplings need to be $\mathcal{O}(1)$ in order to produce a large enough quartic term. It means that unless the triplet VEV is tiny (which requires $\kappa_1$ and $\kappa_2$ to be equal to a very high accuracy), the induced neutrino masses will be too large. This constraint on the $\phi$ VEV is even much stronger than that from the custodial $SU(2)$ violation.

\section{Couplings between SM Fermions and Composite Operators, and Their Peccei-Quinn Charges}\label{PQ charge}

Both SM Yukawa couplings and the Higgs quartic potential from collective symmetry breaking arise from the partial compositeness couplings between the elementary fermions and composite operators. The leading interactions (with $\mathcal{O}(1)$ coupling strength) should respect an approximate $U(1)_{PQ}$ symmetry to avoid a too large quadratic $\tilde{H}_2^\dagger H_1$ term and large FCNCs, so it is convenient to assign the PQ charges to the fermions in classifying the couplings. We will construct a Type-I 2HDM model because of the weaker constraint on the heavy Higgs bosons, and produce both terms needed for the collective quartic Higgs potential.

For the quark sector, we include eight composite operators in $\mathbf{6}$ and $\mathbf{\bar{6}}$ representations of $SU(6)$ with overall PQ charges $r= 0, 1, 2, 3$,
\begin{subequations}
\begin{align}
&\mathbf{6}_{r}&&=\mathbf{2}_{r+1/2}&&\oplus\mathbf{1}_{r}&&\oplus\mathbf{\bar{2}}_{r-1/2}&&\oplus\mathbf{1}_{r}
\\
&\mathbf{\bar{6}}_{r}&&=\mathbf{\bar{2}}_{r-1/2}&&\oplus\mathbf{1}_{r}&&\oplus\mathbf{2}_{r+1/2}&&\oplus\mathbf{1}_{r}
\end{align}
\end{subequations}
Here the subscript denotes the PQ charge instead of the hypercharge.
The $\mathbf{6}$ and $\mathbf{\bar{6}}$ of the same PQ charges create the same resonances which become the quark partners of different flavors. The $U(1)_{PQ}$ charges of the three generations of elementary quarks are shown in Table~\ref{tab:PQ}.
\begin{table}[h]
\centering
\begin{tabular}{||c|c||c|c||c|c||}
\hline
 & $U(1)_{PQ}$ & & $U(1)_{PQ}$ & & $U(1)_{PQ}$ \\ \hline
 $q_{3,L}=(t_L,b_L)^T$ & 1/2 & $t_R$ & 0 & $b_R$ & 1 \\ \hline
 $q_{2,L}=(c_L,s_L)^T$ & 3/2 & $c_R$ & 1 & $s_R$ & 2 \\ \hline
 $q_{1,L}=(u_L,d_L)^T$ & 5/2 & $u_R$ & 3 & $d_R$ & 3 \\ \hline
\end{tabular}
\caption{PQ charges of elementary quarks. The PQ charge of $u_R$ appears out of the pattern. As discussed in the text, the up quark Yukawa coupling comes from the $U(1)_{PQ}$ violating coupling, which also generates the required $\tilde{H}_2^\dagger H_1$ term. \label{tab:PQ}}
\end{table}
The lepton sector can be similarly assigned.

There are some requirements for producing a Type-I 2HDM. First, to generate SM Yukawa couplings, we need to couple one of $q_L$ and $q_R$ to $\mathbf{6}$ and the other to $\mathbf{\bar{6}}$ of the same PQ charge. In addition, each $q_L$ needs to couple to the composite operators at least in two ways in order to generate the up-type and down-type Yukawa couplings with the same Higgs doublet. If $q_L$ had only one coupling to $\mathbf{6}$ (or $\mathbf{\bar{6}}$), the up- and down-type quarks would couple to different Higgs doublets as we discussed in Sec.~\ref{Yukawa}. Once $q_L$ couplings are fixed, the right-handed quark couplings follow directly from the PQ charges (except for the up quark). 
To generate the Higgs quartic term by collective symmetry breaking,
we need to introduce two pairs of couplings between the elementary doublets and the ($\mathbf{6}$, $\mathbf{\bar{6}}$) pairs, with each pair of couplings preserving a different $SU(4)$ symmetry. Finally, we add a $U(1)_{PQ}$ violating $\lambda''_{u_L}$ which serves to generate the mixed Higgs quadratic term in Eq.~\eqref{m12}, and also the up quark Yukawa coupling.

From these requirements, a possible set of couplings between elementary quarks and the composite operators is shown below (in the parentheses after the corresponding composite operators). 
\begin{subequations}
\begin{align}
&\mathbf{6}_{0}&&=\mathbf{2}_{1/2}~(\lambda_{t_L})&&\oplus\mathbf{1}_{0}&&\oplus\mathbf{\bar{2}}_{-1/2}&&\oplus\mathbf{1}_{0}
\\
&\mathbf{\bar{6}}_{0}&&=\mathbf{\bar{2}}_{-1/2}&&\oplus\mathbf{1}_{0}&&\oplus\mathbf{2}_{1/2}&&\oplus\mathbf{1}_{0}~(\lambda_{t_R})
\\
&\mathbf{6}_{1}&&=\mathbf{2}_{3/2}~(\lambda_{c_L})&&\oplus\mathbf{1}_{1}&&\oplus\mathbf{\bar{2}}_{1/2}&&\oplus\mathbf{1}_{1}~(\lambda'_{b_R})
\\
&\mathbf{\bar{6}}_{1}&&=\mathbf{\bar{2}}_{1/2}~(\lambda'_{b_L})&&\oplus\mathbf{1}_{1}&&\oplus\mathbf{2}_{3/2}&&\oplus\mathbf{1}_{1}~(\lambda_{c_R})
\\
&\mathbf{6}_{2}&&=\mathbf{2}_{5/2}&&\oplus\mathbf{1}_{2}&&\oplus\mathbf{\bar{2}}_{3/2}~(\tilde{\lambda}'_{s_L})&&\oplus\mathbf{1}_{2}
\\
&\mathbf{\bar{6}}_{2}&&=\mathbf{\bar{2}}_{3/2}&&\oplus\mathbf{1}_{2}~(\tilde{\lambda}'_{s_R})&&\oplus\mathbf{2}_{5/2}~(\tilde{\lambda}_{u_L})&&\oplus\mathbf{1}_{2}
\\
&\mathbf{6}_{3}&&=\mathbf{2}_{7/2}~(\lambda''_{u_L})&&\oplus\mathbf{1}_{3}&&\oplus\mathbf{\bar{2}}_{5/2}&&\oplus\mathbf{1}_{3}~(\lambda'_{d_R})
\\
&\mathbf{\bar{6}}_{3}&&=\mathbf{\bar{2}}_{5/2}~(\lambda'_{d_L})&&\oplus\mathbf{1}_{3}&&\oplus\mathbf{2}_{7/2}&&\oplus\mathbf{1}_{3}~(\lambda_{u_R})
\end{align}
\end{subequations}
where the subscript of the coupling tells which elementary quark it is coupled to. (The left-handed couplings couple to the whole doublets despite the quark labels.) The SM quark Yukawa couplings are given by
\begin{align}
&y_t\sim\frac{\lambda_{t_L}\lambda_{t_R}}{g_{\psi_0}},&&
y_{b}\sim\frac{\lambda'_{b_L}\lambda'_{b_R}}{g_{\psi_1}},\quad\\
&y_c\sim\frac{\lambda_{c_L}\lambda_{c_R}}{g_{\psi_1}},&&
y_s\sim\frac{\tilde{\lambda}'_{s_L}\tilde{\lambda}'_{s_R}}{g_{\psi_2}}\quad\\
&y_{u}\sim\frac{\lambda''_{u_L}\lambda_{u_R}}{g_{\psi_3}},&&
y_{d}\sim\frac{\lambda'_{d_L}\lambda'_{d_R}}{g_{\psi_3}},
\end{align}
where $g_{\psi_r}$ is the coupling of the strong resonances in $\mathbf{6}_r, \mathbf{\bar{6}}_r$, with their masses given by  $\sim g_{\psi_r} f$. 
To have a relatively light top partner, we should have $g_{\psi_0} \sim 2$, while all other $g_{\psi_r}$'s are expected to be large. The quark flavor mixings (CKM matrix) can be generated by additional $U(1)_{PQ}$ violating couplings which are not shown. These couplings are expected to be small and will not significantly affect the Higgs potential.

For the Higgs quartic term, the combination of $\lambda_{c_L}$ and $\lambda'_{b_L}$ generates one term of the collective symmetry breaking, while the combination of $\tilde{\lambda}'_{s_L}$ and $\tilde{\lambda}_{u_L}$ generates the other. Alternatively, we could also use the lepton sector to generate one of the collective symmetry breaking terms. The quartic coupling is estimated to be 
\begin{equation}
\lambda_{12}=\frac{3}{4\pi^2}\frac{c_{cb}c_{us}\lambda_{c_L}^2\lambda_{b_L}'^2\tilde{\lambda}_{u_L}^2\tilde{\lambda}_{s_L}'^2}
{c_{cb}\lambda_{c_L}^2\lambda_{b_L}'^2+c_{us}\tilde{\lambda}_{u_L}^2\tilde{\lambda}_{s_L}'^2}
\sim \frac{3}{4\pi^2}\lambda_{c_L}^2\lambda_{b_L}'^2  \quad (\mbox{if }  \lambda_{c_L}\lambda_{b_L}' \sim \tilde{\lambda}_{u_L}\tilde{\lambda}_{s_L}',~c_{cb}\sim c_{us}\sim 2). 
\end{equation}
To get a large enough $\lambda_{12}$, these couplings should be quite large ($\gtrsim 1$). The correct SM Yukawa couplings can still be obtained by suitable choices of $\lambda_R$ couplings and $g_{\psi_r}$.
The $\lambda_{u_L}''$ coupling violates the $U(1)_{PQ}$ symmetry as it mixes the $q_{1,L}$ with charge $5/2$ with the composite doublet of charge $7/2$. By combining with $\lambda_{d_L}'$, it will generate a mixing mass term for the two Higgs doublets, 
\begin{equation}
m_{12}^2\sim \frac{3}{8\pi^2}\lambda_{d_L}'\lambda_{u_L}''g_{\psi_3}^2f^2~.
\end{equation}
In this way, all terms required in the Higgs potential for a realistic model can be generated without introducing additional elementary fermions.

\bibliographystyle{jhepbst}
\bibliography{CHM_Ref}{}

\end{document}